\documentclass{aa}  

\usepackage{graphicx}
\usepackage{txfonts}
\usepackage{lipsum}
\usepackage{subcaption}
\usepackage{lscape}       
\usepackage{placeins}           
\usepackage{url}
\usepackage[breaklinks,colorlinks,citecolor=blue,linkcolor=red,urlcolor=magenta]{hyperref}
\usepackage{caption}
\usepackage{longtable}
\usepackage{multicol}
\usepackage{color}
\usepackage{natbib,twoopt}
\usepackage{soul}
\usepackage[dvipsnames]{xcolor}

\let\linenumbers\relax
\let\modulolinenumbers\relax
\let\runningpagewiselinenumbers\relax

\newcommand{\kms}{km~s$^{-1}$}
\newcommand{\whz}{W~$\rm Hz^{-1}$}
\newcommand{\sqdeg}{deg$^2$}
\newcommand{\mjyb}{mJy~beam$^{-1}$}
\newcommand{\ujyb}{$\mu$Jy~beam$^{-1}$}

\def\t     {$\times$ }

\def \deg         {\text{$^{\circ}$}}
\def \arcmin      {\text{$^\prime$}}
\def \arcsec      {\text{$^{\prime\prime}$}}

\begin{document}
\title{Study of giant radio galaxies using spectroscopic observations from the Himalayan Chandra Telescope}
\authorrunning{Sethi \& Dabhade et al.}

\author{Sagar Sethi\thanks{These authors contributed equally to this article.}\fnmsep\thanks{Email: sagar.sethi@doctoral.uj.edu.pl, pratik.dabhade@ncbj.gov.pl}\fnmsep\inst{1,2}
\and Pratik Dabhade\footnotemark[1]\fnmsep\footnotemark[2]\fnmsep\inst{3}
\and K.G. Biju\inst{4}
\and C. S. Stalin\inst{5}
\and Marek Jamrozy\inst{1}
}

\institute{Astronomical Observatory, Jagiellonian University, ul. Orla 171, 30-244 Krakow, Poland
\and{Doctoral School of Exact and Natural Sciences, Jagiellonian University, Krakow, Poland}
\and Astrophysics Division, National Centre for Nuclear Research, Pasteura 7, 02-093 Warsaw, Poland
\and WMO Arts \& Science college, Muttil, Wayanad, Kerala, 673122, India
\and Indian Institute of Astrophysics, Koramangala II Block, Bangalore, 560 034, India}

 
\abstract
{We present the results of spectroscopic observations of host galaxies of eleven candidate giant radio galaxies (GRGs), powered by active galactic nuclei (AGNs), conducted with the 2-m Himalayan Chandra Telescope (HCT). The primary aim of these observations, performed with the Hanle Faint Object Spectrograph Camera (HFOSC), was to secure accurate spectroscopic redshifts, enabling precise calculations of their projected linear sizes. Based on these measurements, we confirm all eleven sources as giants, with linear sizes ranging from 0.7 to 2.9 Mpc, including ten GRGs and one giant radio quasar (GRQ). One of the GRGs shows evidence of a potential AGN jet-driven ionized outflow, extending up to $\sim$12 kpc, which, if confirmed, would represent a rarely observed feature. Two of the confirmed GRGs exceed 2 Mpc in size, which are relatively rare examples of GRG. The redshifts of the host galaxies span \(0.09323 \leq z \leq 0.41134\). Using the obtained spectroscopic data, we characterised their AGN states based on the optical emission line properties. To complement these observations, archival radio and optical survey data were utilised to characterise their large-scale radio morphology and estimate projected linear sizes, arm-length ratios, flux densities, luminosities, and core dominance factors. These results provide new insights into the properties of GRSs and form a critical foundation for further detailed studies of their environments, AGN activity, and evolution using future high-sensitivity optical and radio datasets.}


   \maketitle

\section{Introduction}\label{sec1:intro}
Galaxies with active galactic nuclei (AGNs) at their centres that emit predominantly at radio wavelengths with jets and lobes are called radio galaxies (RGs). They have been extensively studied over the past decades \citep[for review, see;][]{Begelman84,Hardcastle2020,Jet.Saikia.2022}. When powered by quasars, they are referred to as radio quasars (RQs). The radio jets in RGs and RQs can extend from a few kiloparsecs (kpc) to megaparsecs (Mpc). Those with end-to-end projected linear sizes greater than 700 kpc are classified as giant radio galaxies (GRGs) or giant radio quasars (GRQs). Together, they are often referred to as giant radio sources (GRSs); however, throughout this paper, we choose to refer to the population, irrespective of whether they are quasars or not, as GRGs. They represent a rare subset of the RG population \citep[for review, see;][]{GRSREV2023}. GRG research began with the identification of the megaparsec-scale structures of 3C 236 and DA 240, first reported by \citet{Willis1974}. GRGs are valuable laboratories for studying galaxy evolution over cosmic time. However, the reasons behind their rarity compared to smaller RGs and the mechanisms driving their exceptional sizes remain active areas of research \citep[e.g.][]{Kuzmicz2018,sagan1,kuzmicsethi2021,sagan3}. Given their extreme sizes, GRGs are often studied as a distinct population. However, it is possible that they simply represent the upper tail of the continuous size distribution of extended radio galaxies.

GRGs play an essential role in probing large-scale cosmic structures and astrophysical processes. For instance, their interaction with the intergalactic medium (IGM) offers unique insights into AGN feedback, while their large-scale environments can reveal details about cosmic web filaments and voids \citep[e.g.,][]{Oei2023}. Additionally, they can be useful for probing large-scale magnetic field strengths, as demonstrated by \citet{SAGANIV} in supercluster environments. The exceptional potential of GRGs as astrophysical probes has been reviewed in \citet{GRSREV2023}. 

Over the years, advancements in radio astronomy, particularly with surveys like LOFAR Two-metre Sky Survey \citep[LoTSS;][]{lotssdr1.Tim}, have revolutionized the identification of these giants \citep[e.g.,][]{Dabhade2020a,Simonte2022,Oei.LoTSS,2024Mostert}. These surveys, with unprecedented sensitivity and good angular resolution, have revealed numerous diffuse and faint GRGs previously missed in older surveys. Despite these advances, a significant challenge remains: the lack of complementary deep-wide-sky optical spectroscopic surveys to determine precise host galaxy redshifts. Consequently, a substantial fraction of the reported GRG samples rely on photometric redshifts, which, while practical for large surveys, are prone to significant inaccuracies. For instance, \citet{sagan1} achieved $\sim$79\% spectroscopic confirmation in their GRG catalogue compiled from literature, compared to $\sim$63\% in LoTSS-DR1 \citep{Dabhade2020a}, $\sim$7\% in 1059~\sqdeg\, sky area of Rapid ASKAP Continuum Survey (RACS) \citep{Andernach.RACS}, $\sim$50\% in LoTSS-DR2 \citep{Oei.LoTSS}, and $\sim$36\% in LoTSS Deep Fields \citep{Simonte.2024}. This clearly illustrates the urgent need for comprehensive spectroscopic follow-ups.

Below, we highlight several GRG-specific studies that underscore the critical importance of spectroscopic measurements of GRG hosts. Without these measurements, such studies would have been infeasible, and our understanding of GRGs and their unique properties would have remained significantly limited. Recognising the critical limitation posed by the lack of redshift information for GRG candidates, these studies focused on obtaining accurate spectroscopic redshifts to confirm GRG status and enable precise determination of their physical properties. Redshift measurements are indispensable for calculating projected linear sizes, luminosities, and energy densities, which are key to understanding the evolution of RGs and their interaction with their environments.

\begin{itemize}
    \item The studies by \citet{Machalski2004,Machalski2007salt} marked one of the earliest systematic spectroscopic observation programs dedicated to GRGs following the initial groundwork laid by studies such as \citep[e.g.][]{Schoenmakers2001,Lara2001}. \citet{Machalski2007salt} demonstrated how accurate redshifts enable robust comparisons of GRGs across cosmic epochs, revealing key factors driving their extraordinary sizes, such as jet power, environmental density, and source age.

    \item \citet{sagan2iram} used the IRAM 30-m telescope to investigate molecular gas in GRG hosts, requiring precise spectroscopic redshifts to tune instruments to the correct spectral lines, enabling insights into AGN fueling and galaxy evolution.
    
    \item \citet{SAGANIV} used spectroscopic redshifts to classify GRGs into cluster and non-cluster environments, finding 24\% in dense clusters. Accurate redshifts were crucial to avoid misclassification, ensuring reliable analysis of environmental effects on GRG morphology and evolution.
    
    \item Spectroscopic observations of J0644$+$1043 \citep{SGRG.Sethi24} revealed the limitations of photometric redshifts for extragalactic sources near the Galactic plane. The host galaxy's photometric redshift \(z_{\text{phot}} = 0.24\) \citep{Delli} implied a 2.8 Mpc size, but spectroscopy revised it to \(z_{\text{spec}} = 0.0488\), reducing the size to 0.71 Mpc.

\end{itemize}

These studies collectively highlight the importance of spectroscopic redshifts in advancing GRG research, just as they are essential for other extragalactic objects. From enabling robust environmental classifications and multi-wavelength analyses to correcting critical inaccuracies in photometric data, spectroscopic campaigns remain essential for advancing our understanding of GRGs and their role in the cosmic ecosystem.

The Sloan Digital Sky Survey \citep[SDSS;][]{SDSS.York00} revolutionized GRG research by providing large-scale spectroscopic data \citep[e.g.,][]{d17}, enabling robust redshift measurements for thousands of sources. However, SDSS is constrained by its focus on the northern hemisphere and its limited sensitivity to redshifts beyond $\sim 0.7$ \citep{SDSS.z.limit}. As a result, a substantial fraction of GRG candidates, particularly those in the southern sky or at higher redshifts, remain unconfirmed, underscoring the continued necessity of dedicated spectroscopic campaigns.

To address specific problems posed by GRGs, the project SAGAN\footnote{\url{https://sites.google.com/site/anantasakyatta/sagan}} (Search and Analysis of GRGs with Associated Nuclei) was initiated \citep{d17}. Project SAGAN has systematically identified and analyzed GRGs by integrating large-scale radio surveys with optical spectroscopic data \citep[e.g.,][]{sagan1}. This effort led to the creation of the first large GRG catalogue \citep{Dabhade2020a}, which significantly advanced GRG studies by reporting more than twice the population of giants compared to the initial census of $\sim$300 sources by \citet{d17}.

This study extends the legacy of SAGAN by targeting GRG candidates lacking spectroscopic redshifts in public archives. We conducted observations with the 2-m Himalayan Chandra Telescope\footnote{\url{https://www.iiap.res.in/centers/iao/facilities/hct/}} (HCT), obtaining precise redshift measurements for eleven host galaxies.

In the current paper, we present the results of our optical spectroscopic observations and their derived optical and radio properties. In this process, we report accurate redshifts of 10 GRGs and 1 GRQ, of which three sources are being reported as GRG for the first time. We discuss our sample characteristics, observation details and data reduction procedures in Sec.~\ref{sec2:sample} and available archival data in Sec.~\ref{sec3:archival}. In Sec.~\ref{sec4:obs_results}, we presented results from our spectroscopy and radio analysis. Notes on individual GRGs are presented in Sec.~\ref{sec5:new_grgs}, followed by a summary and discussion in Sec.~\ref{sec6:summary}.

Throughout the paper, we adopt the flat $\Lambda$CDM cosmological model based on the Planck results \citep[$\rm H_0= 67.8$\, \kms Mpc$\rm ^{-1}$, $\rm \Omega_m= 0.308$;][]{Plank2016}. All images are in the J2000.0 coordinate system. We use the convention S$\rm_{\nu}\propto \nu^{-\alpha}$, where S$_{\nu}$ is the flux density at frequency $\nu$ and $\alpha$ is the spectral index. Forbidden lines are denoted in square brackets throughout the paper and in the images. All identified skylines in the spectra are marked as a circled cross mark on the respective spectra.

\section{Sample, observations, and analysis}\label{sec2:sample}
\subsection{Sample}
\begin{figure*}
    \centering
    \includegraphics[scale=0.06]{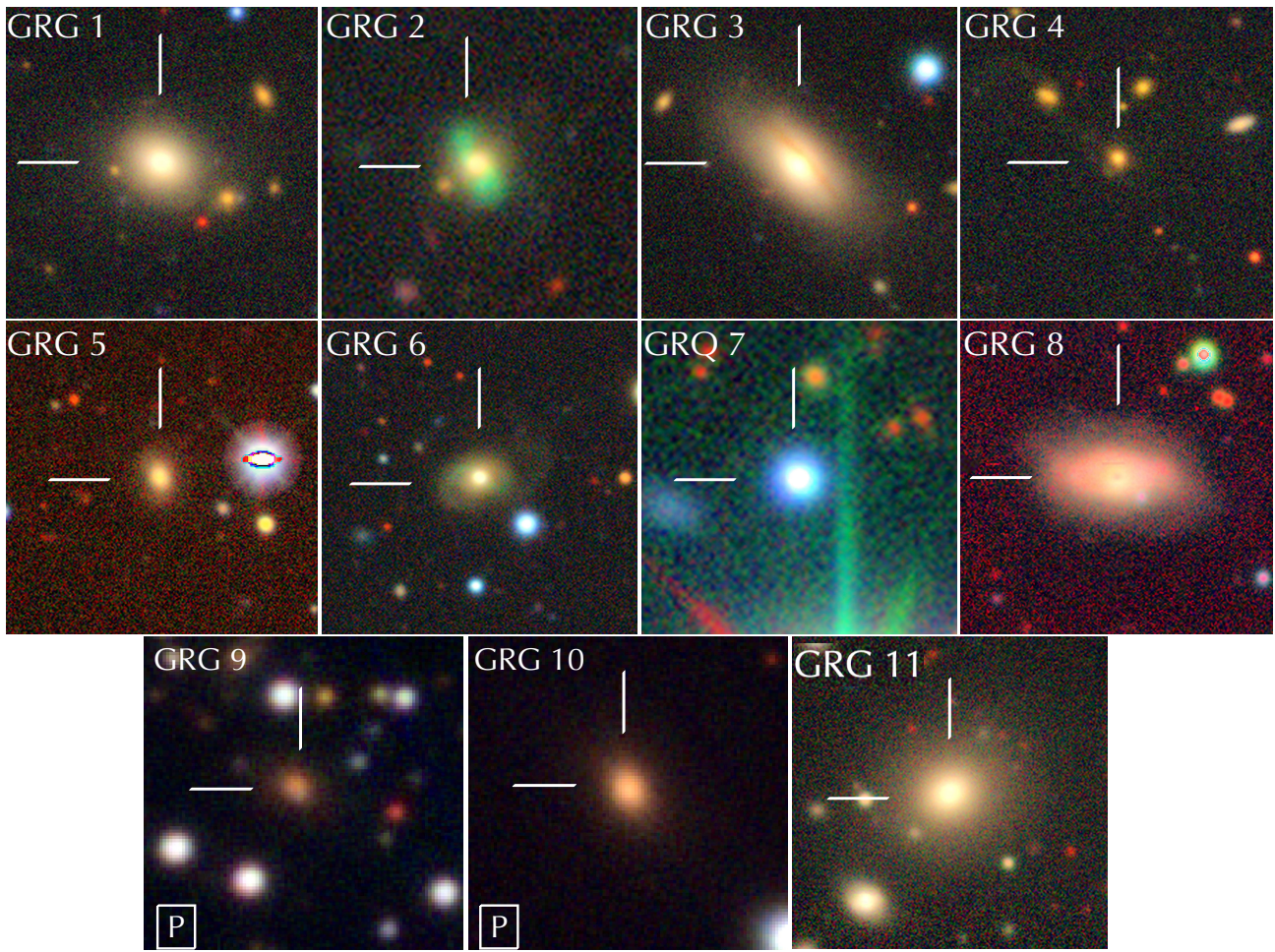}
    \caption{Montage of optical colour images of host galaxies at centre (highlighted using two white marker lines) of 10 GRGs and 1 GRQ from DESCaLS-DR9 or PanSTARRS-DR1 (marked with P in a box) image. More details can be found in Sec.\,\ref{subsec3:opdata}.}
    \label{fig:optical-montage}
\end{figure*}

As a result of our extensive search for GRGs in the available large sky area radio surveys, we have found several extended radio sources with low radio surface brightness. While we successfully identified their optical host galaxies using optical survey data (e.g., SDSS), many lacked spectroscopic redshift information. Based on their angular sizes and either photometric redshifts or the faintness of their host galaxies, we identified these sources as GRG candidates. In this sample of GRG candidates, there are instances where the optical host galaxies of radio sources are found to reside in crowded systems on the plane of the sky, either in the galaxy group or areas heavily populated by foreground objects from our own galaxy (for example, see GRG9 in our sample; refer to Fig.\,\ref{fig:optical-montage}). Also, in some cases, the optical host galaxies appear faint in the optical band or are located very close to foreground objects within our galaxy (for example, see GRQ7 in Fig.\,\ref{fig:optical-montage}). Hence, possibly due to the above-mentioned factors, these sources were likely not included in large spectroscopic surveys, making dedicated and careful spectroscopic observations essential for their study.

Our sample of low radio surface brightness GRG candidates, lacking spectroscopic redshift measurements, was filtered for observations with the HCT based on the following criteria: (i) sources within the sky observable by the HCT, (ii) host galaxies with apparent magnitudes brighter than  m$\rm _{r}=20$, and (iii) sources accessible during the allocated observing runs based on their LST. Our final sample consists of 11 sources, with host galaxies having $r$-band magnitudes spanning $13.2 \leq m_\mathrm{r} \leq 19.5$, ensuring they are within the brightness limits suitable for spectroscopic observations with the HCT.

\setlength{\tabcolsep}{3pt}
\begin{table*}
\centering
\caption{Obervation log: Col(1): Serial number, Col(2): Object name, Col(3) and Col(4): Right ascension (R.A) in HMS and declination (Dec) in DMS, respectively, for the host galaxies of the GRGs, Col(5): Apparent $r$-band magnitude, Col(6): Observation date, Col(7): Grism used during the observation, Col(8): Total exposure time of the observation for each object, Col(9): References for the first time reporting of each GRG - 1 for \citet{sagan1}, 2 for new GRGs reported in this paper, 3 for \citet{d17}, 4 for \citet{Machalski2007salt} and 5 for \citet{Oei.LoTSS}, Col(10): $z \rm_{photo}$ values of the GRGs, superscript `s' and `d' indicates $z \rm_{photo}$ obtained from SDSS and DESCaLS \citep{DESCal.Duncan22} respectively.}
\begin{tabular}{cccccccccc}
\hline
GRG No & Name &R.A &Dec & m$\rm_{r}$ &Observing date &Grism &Total exp (mins) & Ref & $z \rm_{photo}$\\
(1) &(2) &(3) &(4) &(5) &(6) &(7) &(8) &(9) &(10)\\
\hline
1 & J0125$+$0703 & 01 25 32.17 & $+$07 03 37.10 &16.0 & 10$^{\rm th}$ Jan 2016 & 7 & 75 & 1 & 0.098$\rm ^s$\\
2 & J0151$-$1112 & 01 51 24.94 & $-$11 12 00.58 &16.4 & 28$^{\rm th}$ Nov 2016 & 8 & 30 & 2 & 0.499$\rm ^d$\\
3 & J0235$+$1011 & 02 35 12.71 & $+$10 11 46.72 &15.8 & 10$^{\rm th}$ Jan 2016 & 7 & 75 & 5 & 0.061$\rm ^d$\\
4 & J0429$+$0033 & 04 29 25.85 & $+$00 33 04.81 & 19.5 & 10$^{\rm th}$ Jan 2016 & 5 & 90 & 3 & 0.468$\rm ^s$\\
5 & J0754$+$2324 & 07 54 41.67 & $+$23 24 22.86 &18.0 & 9$^{\rm th}$ Jan 2016 & 7 & 75 & 2 & 0.133$\rm ^s$\\
6 & J0824$+$0140 & 08 24 18.16 & $+$01 40 40.21 & 17.5 & 9$^{\rm th}$ Jan 2016 & 7 & 60 & 4 & 0.204$\rm ^d$\\
7 & J0847$+$3831 & 08 47 46.06 & $+$38 31 39.33 & 16.6 & 28$^{\rm th}$ Nov 2016 & 8 & 60 & 1 & 0.450$\rm ^d$\\
8 & J1908$+$6957 & 19 08 07.41 & $+$69 57 59.40 & 13.2 & 4$^{\rm th}$ June 2016 & 8 & 15 & 5 & 0.067$\rm ^d$\\
9 & J1930$+$2803 & 19 30 11.80 & $+$28 03 30.11 & 18.0 & 5$^{\rm th}$ June 2016 & 8 & 60 & 2 & -\\
10 & J2059$+$2434 & 20 59 39.82 & $+$24 34 23.94 & 16.3 & 27$^{\rm th}$ Nov 2016 & 8 & 30 & 3 & 0.116$\rm ^s$\\
11 & J2250$+$2844 & 22 50 39.15 & $+$28 44 45.51 & 14.4 & 5$^{\rm th}$ June 2016 & 8 & 27 & 3 & 0.097$\rm ^s$\\
\hline
\end{tabular}
\label{tab:1_sample}
\end{table*}

\subsection{Observations with HCT}\label{subsec:2_observation}
The HCT, with its flexibility for targeted observations and spectroscopic capabilities, is particularly suited for studying host galaxies of GRGs that are often overlooked by large spectroscopic surveys like SDSS or DECam Legacy Survey \citep[DECaLS;][]{DESCal}. Its ability to focus on specific targets, especially in crowded or low-surface-brightness regions, enables the acquisition of precise redshifts for sources that might otherwise remain unclassified. 

Optical long-slit spectroscopy data were observed in three observation cycles during nights of 9$^{\rm th}$-10$^{\rm th}$ in January, 4$^{\rm th}$-5$^{\rm th}$ in June and 27$^{\rm th}$-28$^{\rm th}$ in November 2016 (for details please refer to the Tab.\,\ref{tab:1_sample}) with the Hanle Faint Object Spectrograph and Camera (HFOSC\footnote{\url{https://www.iiap.res.in/centers/iao/facilities/hct/hfosc/}}). HFOSC is a spectrograph mounted on the 2-m HCT at the Indian Astronomical Observatory (IAO) in Hanle (longitude: 78\deg 57\arcsec 51\arcmin\,E, latitude: 32\deg 46\arcmin 46\arcsec\,N, altitude $\sim$\,4500 m; \citealt{prabhuanu10}). HFOSC employs a SITe ST-002 2K\,$\times$\,4K, thinned, back-illuminated CCD with 15$\muup$m pixels. The spatial sampling scale at the detector is 870 nm per arcsecond, giving a field of view of about 10\arcmin\,on the side. We used HFOSC \citep{iao02,iao14} grisms 5, 7, and 8 with different slit options. Grism 5 has a spectral resolution ($\frac{\lambda}{\delta\lambda}$) of 870 with a wavelength coverage of 5200-10300 ($\AA$).  Grism 7 offers a wavelength coverage of 3800-6840 ($\AA$) and a spectral resolution ($\frac{\lambda}{\delta\lambda}$) of 1330. For grism 8, the wavelength coverage is 5800-8350 ($\AA$) and the spectral resolution ($\frac{\lambda}{\delta\lambda}$) is 2190. Standard spectroscopy observation techniques were used during each observation. We obtained flat, bias, spectroscopic lamps (HFOSC uses Fe-Ne and Fe-Ar spectroscopic lamps) and target frames. The seeing conditions during observations were typically between 1.5\arcsec\, and 2.3\arcsec\,. Except for GRG8 (the $r$-band magnitude, m$\rm _{r}$ is 13.2), all other sources were observed for $\geq$ 1800 seconds.

\subsection{Analysis: Spectroscopic calibration and data reduction}\label{subsec:2_data_reduction}\
We performed the data reduction and calibration using the NOAO Image Reduction and Analysis Facility \citep[IRAF;][]{IRAF86,IRAF93} package. The data reduction is divided into a pre-processing step and a spectrum reduction step. We applied bias and flat-field corrections to all images in the pre-processing step. Cosmic ray hits were removed using the \texttt{crmedian} task in the Cosmic Ray Removal Utility Package \texttt{crutil}. Dark correction was not required because the CCD was cooled with liquid nitrogen and the dark current was negligible for 45-minute exposures. We created bias-subtracted flat-field normalised images using the \texttt{ccdproc} task in the \texttt{ccdred} package.

For the spectrum reduction step, we used the package \texttt{specred}. It includes tasks \texttt{apall, identify, refspec, dispcor}. We extracted the spectra of the object and lamp with task \texttt{apall}, using a variance-weighted extraction method compared to the normal method. The wavelength calibration of the object spectra was performed with the standard spectra of the iron-neon (Fe-Ne) or iron-argon (Fe-Ar) lamps in the task \texttt{identify, refspec, dispcor}. We calculated the Gaussian centre for the emission lines in the object and for lamp spectra using Gaussian fitting routines \citep[using Astropy;][]{astropy} for the most prominent emission and absorption lines. The instrumental broadening (FWHM$\rm_{instr}$) of grism 5, 7 and 8 are 6.27 ($\AA$), 4.85 ($\AA$) and 6.78 ($\AA$), respectively. We applied the International Astronomical Union (IAU) standard method to convert air wavelengths to vacuum wavelengths, as described in \citet{Vac2air-M91}, during our redshift calculations. 

\section{Archival multi-wavelength data}\label{sec3:archival}
\subsection{Optical data}\label{subsec3:opdata}
The host galaxies of all GRGs are detected in all filters ($u$, $g$, $r$, $i$, and $z$) of the Panoramic Survey Telescope and Rapid Response System data release-1 \citep[PanSTARRS-DR1;][]{Pan-STARRS.DR1.2016}. The host galaxies of six GRGs and their respective $z \rm_{photo}$ are available in SDSS DR16. Nine GRGs are also covered by the DECaLS-DR9 and their respective $z \rm_{photo}$ are also available in \cite{DESCal.Duncan22}. A montage of colour images from DECaLS-DR9 or PanSTARRS-DR1 of these host galaxies is presented in Fig.\,\ref{fig:optical-montage}. Notes on their optical properties are included in individual notes in later sections.

\subsection{Radio data}\label{subsec3:radiodata}
All GRGs in our sample are available in the NRAO VLA Sky Survey \citep[NVSS;][]{nvss} and the VLA Sky Survey \citep[VLASS;][]{VLASS}. Two GRGs (GRQ7 and GRG12) are also well detected in the LoTSS-DR2. Details on GRGs detected in LoTSS are discussed in a later section. We use VLASS to reliably identify the radio core and host galaxy coincidence, as well as the core strength. We found that the compact radio core for each GRG is available and coincides with the spectroscopic observation target. 

\section{Results}\label{sec4:obs_results}
\subsection{Spectroscopic redshifts}\label{subsec4:spec_redshift}
Several prominent emission and absorption lines along with some telluric lines were identified in our spectra. Among the detected lines were the H$\beta$, H$\alpha$, [O\,{\sc iii}] lines, [N\,{\sc ii}] lines, the Na doublet and the Mg triplet. On the basis of the wavelength shift, the redshifts were estimated for individual lines (listed in Tab.\,\ref{tab:2_spectra_lines} of Appendix ~\ref{sec:app_line}). The final redshift for each host galaxy was determined by averaging the redshifts calculated from each identified emission line. This final redshift, along with its associated error, is presented in Col. 6 of Tab.\,\ref{tab:2_spectra_lines}. Redshift errors were calculated using the corrected full width at half maximum (FWHM) of observed spectral lines, accounting for instrumental broadening. This multi-line approach ensures robust and precise redshift estimates, reducing uncertainties arising from individual line measurements. The redshifts obtained for our GRG sample ranged from 0.09323 to 0.41134. All wavelength-calibrated spectra in the observed frame are presented in the Appendix ~\ref{sec:app} (Fig.\,\ref{fig:0125} to \ref{fig:2250}). Using these redshift measurements we estimated the projected linear size of the GRGs, as listed in Tab.\,\ref{tab:2radio}. These optical spectra and precise redshift measurements not only confirm the giant nature of these radio sources but also lay the groundwork for studying their environments and AGN accretion dynamics.

\subsection{Excitation state of AGNs powering GRGs}\label{subsec4:ex}
The emission and absorption lines in the optical spectrum of an AGN provide critical insights into the physical and chemical properties of the narrow and broad line regions, revealing the signatures of elements and ionisation processes within these regions \citep{netzer_2013}. The presence of prominent narrow and broad features in the optical spectra resulted from the excited emission lines of the AGNs' nuclear activity, indicating the AGN's excitation state. Therefore, prominent line features in the optical spectrum distinguish high-excitation AGNs from low-excitation AGNs. The host of RGs that show no broad or strong narrow emission lines are classified as low-excitation radio galaxies (LERG). On the other hand, RG hosts with broad and narrow emission lines are classified as high-excitation radio galaxies (HERG). Understanding the excitation state of AGNs is critical for probing the accretion dynamics and feedback mechanisms that drive the evolution of GRGs. The excitation state directly impacts the energy output and interaction of jets with their environments, shaping the morphology and properties of GRGs.

To classify the excitation state of the AGNs in the GRGs from our sample, we adopted the criteria outlined by \citet{Best.Heckman.opt.excitation}, where AGNs with an [O\,{\sc iii}] ($\lambda$5007\AA) line equivalent width $>$ 5 \AA\ are classified as HERGs, while those with lower equivalent widths are categorized as LERGs. Based on this criterion, we classified 6 out of 11 sources as LERGs, while the remaining are categorised as HERGs. Our classification from spectroscopic data was also supported with a mid-infrared diagnostic scheme \citep[e.g.][]{WISE.color.plot.Gurkan,sagan1} using four mid-IR bands of Wide-field Infrared Survey Explorer \citep[WISE;][]{wise}. HERGs can be further subdivided into narrow-line radio galaxies (NLRGs) and broad-line radio galaxies (BLRGs) \citep{WISE.color.plot.Gurkan,Hardcastle2020}. The spectra of NLRGs are similar to those of Seyfert 2, which have permitted and forbidden emission lines of narrow widths. On the other hand, BLRGs have emission lines with very broad widths ($>$2000 km~s$^{-1}$). Hence, based on the above criteria, we have classified four sources from our sample to be NLRGs (GRG2, GRG4, GRG6, and GRG9). The optical spectra of these sources contain a prominent [O\,{\sc iii}]($\lambda$5007\AA) line and their equivalent width of [O\,{\sc iii}] ($\lambda$5007\AA) are larger than $>$5 \AA. The HERG or NLRG nature of these sources is also confirmed using the WISE colour-colour scheme. The optical spectrum for GRQ7 is also present in SDSS-DR16 \citep{SDSSDR16}, our observations with HCT were carried out in 2016, before the release of SDSS-DR16 data. The redshift and classification (quasar) of SDSS-DR16 and our analysis are in agreement. GRG5 is classified as a LERG in the WISE colour-colour plot and has $<$5 \AA~ equivalent width of [O\,{\sc iii}] ($\lambda$5007\AA). With no prominent emission lines, GRG1, GRG3, GRG8, GRG10, and GRG11 are classified as LERGs. All our classification can be seen in the 7$^{\rm th}$ column of Tab.\,\ref{tab:2_spectra_lines}. The dominance of the LERG population in our sample aligns with the findings of \citet{sagan1}, who observed a similar trend in their study with a larger GRG sample and reflects the broader pattern in GRG populations, where radiatively inefficient accretion modes are more common. Overall, these classifications provide crucial insights into the accretion states and feedback mechanisms of GRGs, offering a foundation for understanding their evolution and role in shaping large-scale structures.

\subsection{Radio properties}\label{sec4:radioproperty}
The radio properties of GRGs, such as their morphology, angular sizes, and core dominance factors, are essential for understanding their jet dynamics, growth mechanisms, and interactions with the IGM. All GRGs in our sample are classified as Fanaroff–Riley type II \citep[FR-II;][]{FR-type} or edge-brightened RGs using the NVSS or LoTSS DR2 radio maps. Our sample includes three newly identified GRGs (GRG2, GRG5, and GRG9; marked in bold font in Tab.\,\ref{tab:2radio}), with sizes ranging up to 2.92 Mpc. These rare, large-scale systems provide a unique opportunity to study the extreme ends of GRG size distributions and their physical drivers. The montage of NVSS maps of the three new GRGs is presented in Fig.\,\ref{fig:montage}, with the host galaxy marked as a red circle. The angular size refers to the largest angular distance between two points on the plane of the sky. For FR-II sources, this is measured as the distance between the two hotspot peaks. For all sources, we have measured their angular sizes using the best available radio maps and estimated their projected linear sizes (see Tab.\,\ref{tab:2radio}).

The total 1.4~GHz integrated flux density of the GRGs was estimated using the Common Astronomy Software Applications \citep[\texttt{CASA};][]{casa-team22} task \texttt{imview} by manually selecting the radio emission regions from the NVSS maps. Using the total integrated flux, the total radio power of each source was estimated at 1.4~GHz, assuming a spectral index of 0.75. The corresponding total radio power for our sample varies from 10$^{24}$ to 10$^{25}$ \whz. The core flux densities were obtained from the 3~GHz VLASS \citep{VLASS.Gordon}. Using the core flux, the core radio power of each source was estimated at 3~GHz, assuming a spectral index of 0.5. The corresponding core radio power for our sample varies from 10$^{23}$ to 10$^{24}$ \whz. All radio properties discussed above are listed in Tab.\,\ref{tab:2radio}.

\subsubsection{Core dominance factor:}
We calculated the core dominance factor (CDF) for our sample (see the 8$^{\rm th}$ column of Tab.\,\ref{tab:2radio}) following the methodology from \citet{sagan3}. The CDF is defined as the ratio of the radio core flux density to the extended flux density in the rest frame at 5\,GHz. It is commonly used as an indicator of the orientation of the jet axis relative to the line of sight. The CDF for our sample ranges from 0.03 to 2.18, which falls within the range reported by \citet[][]{GRSREV2023} for a large GRG sample \citep[refer Fig.\,8 in][]{GRSREV2023}.

\subsubsection{Asymmetry in GRGs and its implications:}
GRGs serve as valuable tools for probing the megaparsec-scale environment, offering insights into the dynamics of radio jets and their interaction with the surrounding IGM. One key property of GRGs is their structural asymmetry, often quantified through parameters like the arm-length ratio (ALR) or R$_\theta$, which measures the relative lengths of the jets from the core to the hotspots. The IGM density plays a significant role in influencing jet propagation, and an anisotropic IGM density can lead to arm-length asymmetry, as demonstrated in previous studies \citep[e.g.,][]{arm-length-McCarthy,arm.length.priya.GRG}. Another contributing factor to the observed asymmetry could be orientation effects, which depend on light-travel time and the relative alignment of the jets with the observer's line of sight \citep[e.g.,][]{arm-length.Rudnick}. While previous works have explored the impact of such asymmetries in RGs \citep[e.g.,][]{Gopal_ALR}, a detailed and systematic analysis of these properties within a larger, homogeneous sample of GRGs and RGs remains to be carried out. Such studies are crucial for understanding the physical processes governing jet propagation and the role of environmental interactions and orientation in shaping the observed morphologies of GRGs.

For our sample, we calculated R$_\theta$, which varies from 1.12 to 1.59 (see the 9$^{\rm th}$ column of Tab.\,\ref{tab:2radio}). The mean and median values of R$_\theta$ for our sample are 1.33 and 1.25, respectively. Notably, the median R$_\theta$ value aligns well with results from previous studies, such as \citet{Ishwara1999}, who reported a median R$_\theta$ of 1.39 for FR-II GRGs, \citet{Schoenmakers2000} with R$_\theta$ = 1.28, and \citet{Lara2001} with R$_\theta$ = 1.26. These consistent values suggest that the asymmetries observed in our sample are typical of large-scale radio sources and reinforce the role of environmental and intrinsic factors in shaping jet propagation. Interestingly, these median values for GRGs differ significantly from those of normal-sized RGs. For instance, the 3CRR catalogue reported a median R$_\theta$=1.19. It is important to note that the analysis of asymmetry trends in GRGs is subject to selection effects, as sources with larger angular sizes, fainter cores, or highly asymmetric morphologies are less likely to be reliably identified. These biases should be considered when interpreting statistical trends in GRG structural asymmetry.

\begin{table*}[htb]
\centering
\setlength{\tabcolsep}{5pt}
\caption{Radio Properties. Col(1): Name (all newly reported GRGs are marked in bold fonts) Col(2): Angular size (in arc minutes) calculated from the NVSS maps. We choose a size in between the two peak values of hotspots, Col(3): Size of the source in megaparsec, Col(4): Total integrated flux at 1.4~GHz from NVSS map in mJy, Col(5): Total radio power at 1.4~GHz from the total integrated flux in \whz we assumed a spectral index of 0.75, Col(6): 3~GHz VLASS core flux obtained from \citet{VLASS.Gordon}, Col(7): Core radio power at 3~GHz from core flux in \whz we assumed a spectral index of 0.5, Col(8): The core dominance factor, Col(9): Arm-length ratio (R$_{\rm \theta}$).} 
\label{tab:2radio}
\begin{tabular}{ccccccccc}
\hline

Name &Size &Size &$\rm S_{total}^{1.4~GHz}$ &$\rm P_{total}^{1.4 GHz}$ &$\rm S_{core}^{3~GHz}$ &$\rm P_{core}^{3~GHz}$ &CDF$_{\rm 5GHz}$ & R$_{\rm \theta}$ \\
&(\arcmin) & (Mpc) &(mJy) & (10$^{25}$\whz) &(mJy) & (10$^{24}$\whz) & & \\
(1) &(2) &(3) &(4) &(5) &(6) &(7) &(8) &(9) \\
\hline
J0125+0703 & 8.8 & 1.9 & 108.1$\pm$3.2 & 1.55 & 1.6$\pm$0.2 & 0.22 &0.03 &1.12 \\
\textbf{J0151-1112} & 11.0 & 2.8 & 122.0$\pm$3.0 & 2.70 & 13.1$\pm$0.3 & 2.73 &0.36 & 1.23  \\
J0235+1011 & 14.3 & 2.9 & 151.7$\pm$4.2 & 1.76 & 4.7$\pm$0.2 & 0.53 &0.21 &1.53  \\
J0429+0033 & 5.1 & 1.7 & 121.0$\pm$2.6 & 7.08 & 9.5$\pm$0.2 & 5.10 &0.57 &1.22  \\
\textbf{J0754+2324} & 5.1 & 1.0 & 129.0$\pm$2.2 & 1.44 & - & - &- & 1.25  \\
J0824+0140 & 4.5 & 1.1 & 57.0$\pm$2.7 & 1.02 & 9.1$\pm$0.3 & 1.54 & 0.26 & 1.23  \\
J0847+3831 & 3.0 & 0.9 & 91.0$\pm$1.8 & 2.87 & 10.8$\pm$0.3 & 3.18 &0.10 & 1.47  \\
J1908+6957 & 8.5 & 1.9 & 43.5$\pm$2.5 & 0.61 & 4.3$\pm$0.3 & 0.57 &0.50 &1.17 \\
\textbf{J1930+2803} & 13.1 & 1.4 & 234.0$\pm$4.2 & 0.53 & 25.3$\pm$0.5 & 0.56 &0.31 &1.33 \\
J2059+2434 & 6.7 & 0.7 & 150.0$\pm$2.7 & 0.34 & 12.4$\pm$0.3 & 0.28 &0.46 &1.59 \\
J2250+2844 & 8.2 & 0.9 & 125.0$\pm$2.4 & 0.31 & 52.6$\pm$1.1 & 1.26 &2.18 &1.53 \\
\hline
\end{tabular}
\end{table*}

\begin{figure*}
    \centering
    \includegraphics[scale=0.35]{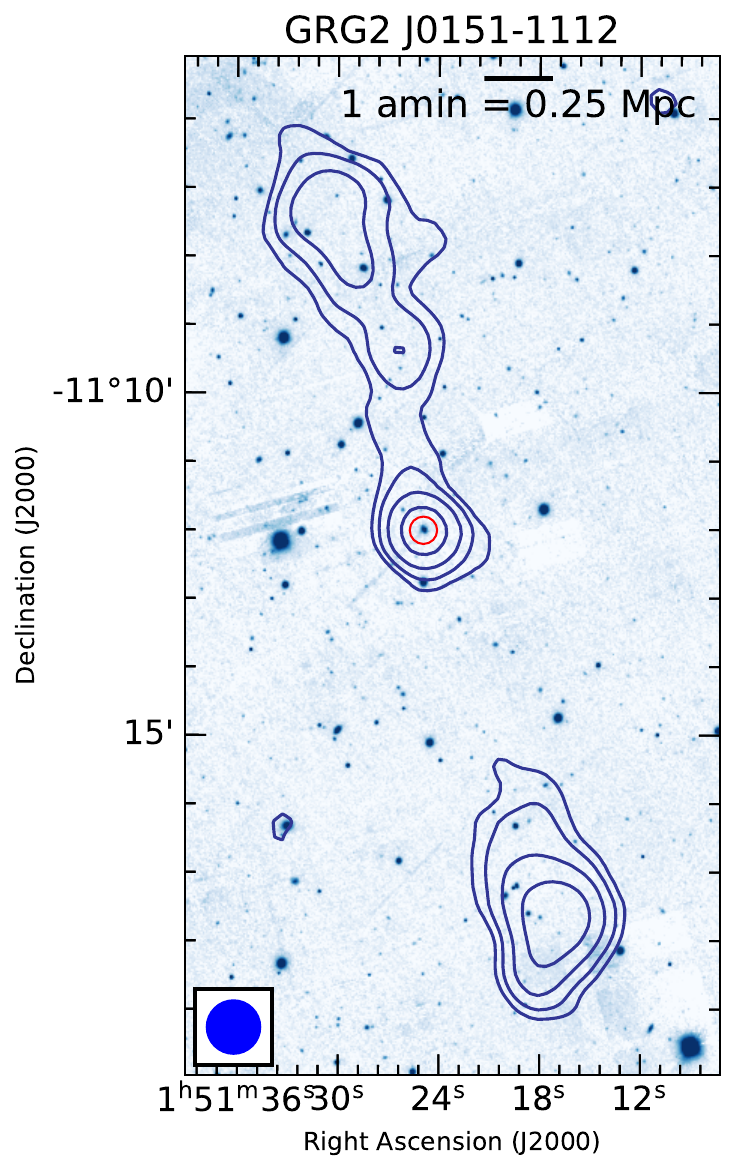}
    \includegraphics[scale=0.35]{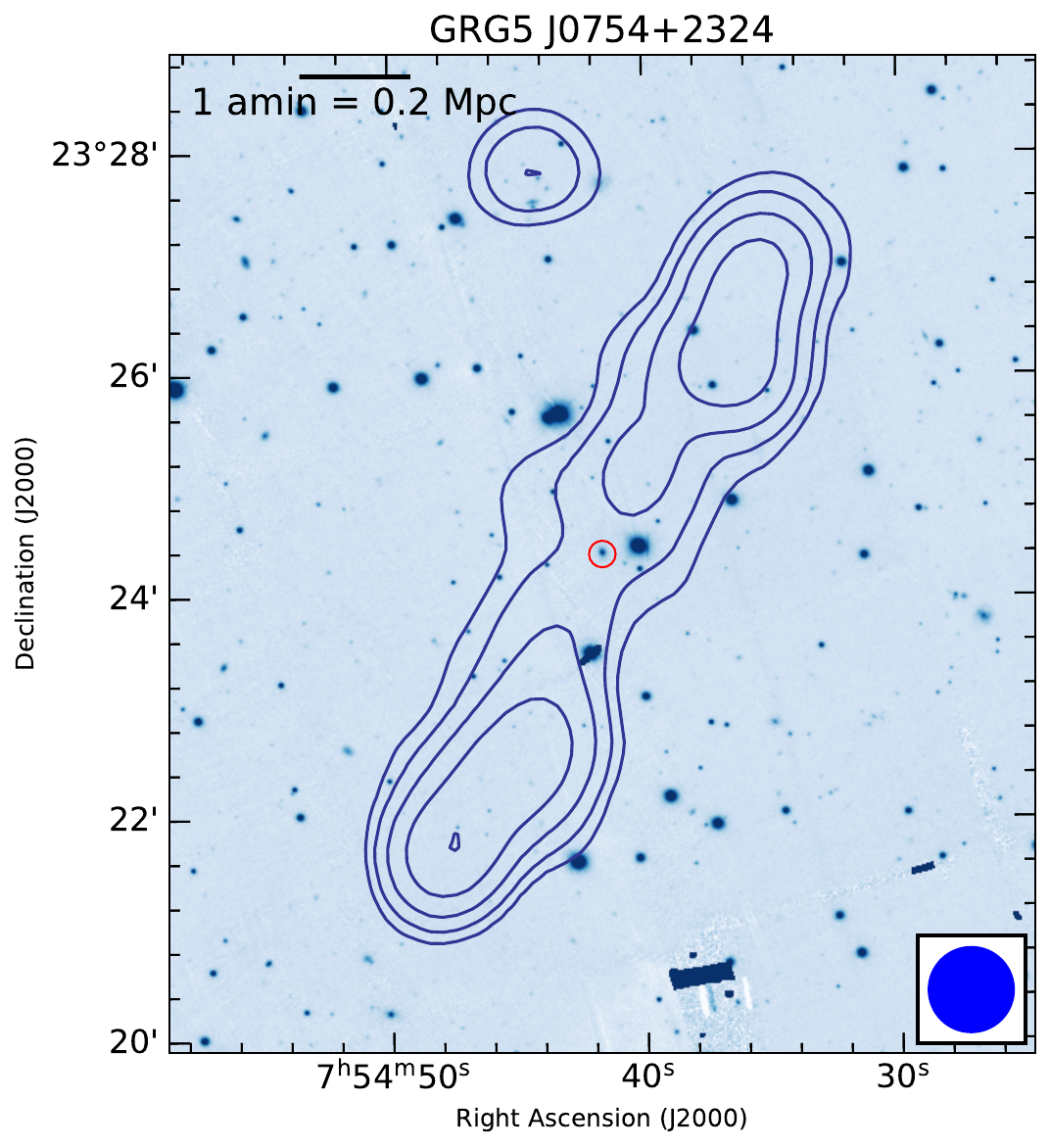}
    \includegraphics[scale=0.35]{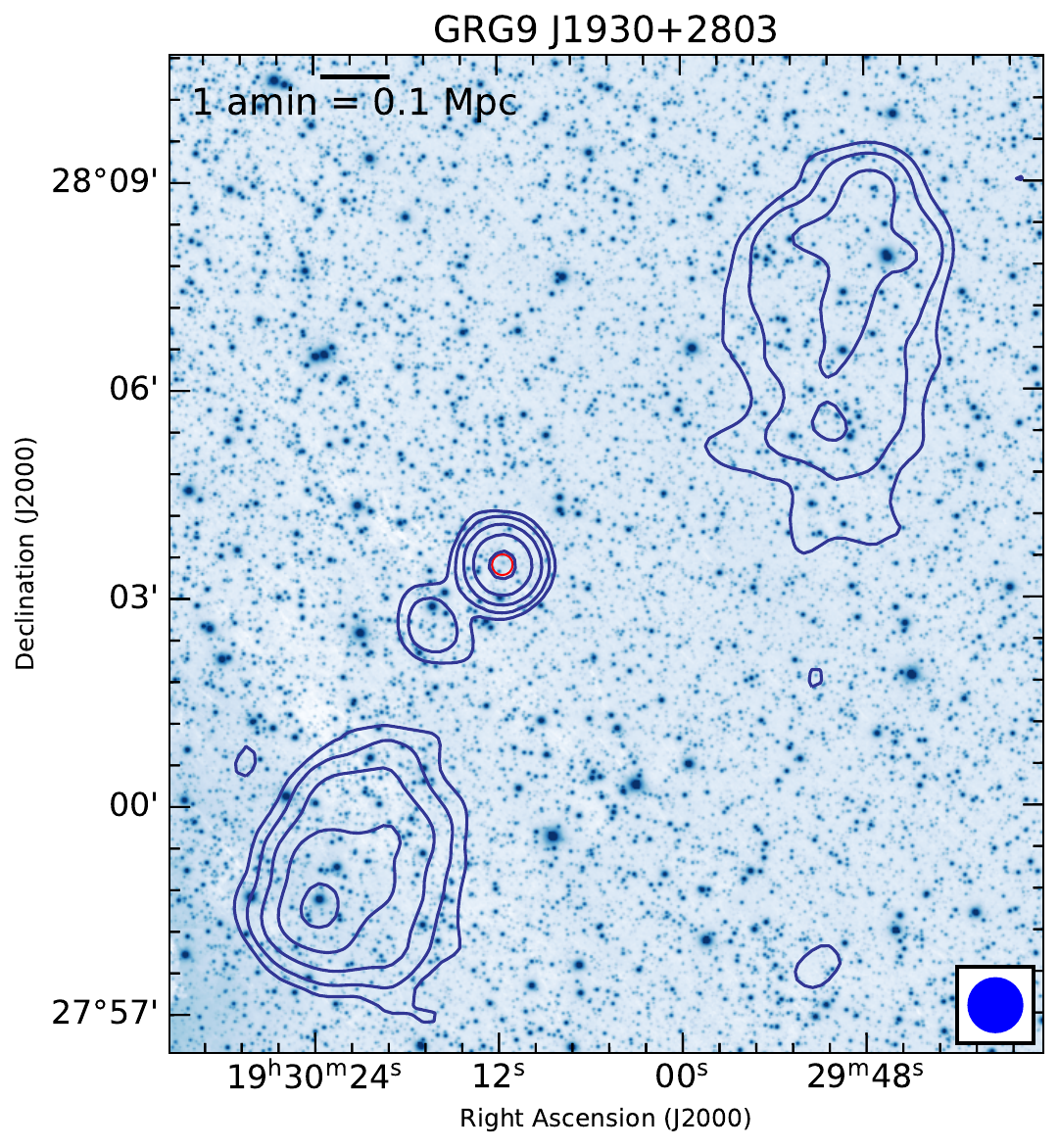}
    \caption{Montage of new GRGs: Each GRG shown in the NVSS contours overlay on the PanSTARRS-DR1 $r$-band map. The Pan-STARRS images are presented in colour map while the NVSS contours are plotted at 1.35\,\mjyb\,$\times\,2^n$ (n = 0,1,2,3 ...) for all the maps. The beam size (in blue colour) is placed in the bottom right or left corner, and the host galaxy position is marked as a red circle for each map. The angular length scale is given in the corners of the relevant figures.}
    \label{fig:montage}
\end{figure*}
\section{Notes on individual GRGs}\label{sec5:new_grgs}
This section presents brief notes on selected sources to highlight their interesting aspects.

\subsection{GRG2 (J0151$-$1112)}
GRG2, the second-largest source in our sample, extends from the northeast (NE) to the southwest (SW) direction, with a projected linear size of 2.79 Mpc. High-resolution images from DECaLS-DR9 (Fig.\,\ref{fig:optical-montage}) reveal a distinct green optical structure emerging from the host galaxy, prominently detected in the g- and $r$-bands. This green emission reflects the strong [O\,{\sc iii}] emission line observed in the optical spectrum of the host galaxy (Fig.\,\ref{fig:0151}), indicating an ionized [O\,{\sc iii}] outflow likely driven by the AGN jet. The [O\,{\sc iii}] emission extends symmetrically to a projected size of $\sim$12 kpc in both directions and aligns with the radio lobe direction from the core, as seen in the VLASS map. This alignment strongly supports the interpretation of a jet-driven outflow, where the jets ionize the surrounding medium, producing the observed [O\,{\sc iii}] emission. Such outflows are critical signatures of AGN feedback, where jet-ISM interactions drive ionization and influence the host galaxy's environment. Further high-resolution spectroscopy and radio observations are necessary to explore the dynamics and impact of this outflow in greater detail. If confirmed, it would represent one of the rare AGN jet-driven ionized outflows with a size exceeding 10 kpc \citep[e.g.,][]{2019Jarvis}.

\subsection{GRG3 (J0235$+$1011)}
GRG3 is the largest source in our reporting sample with a projected linear size of 2.92 Mpc. GRGs of sizes $\sim$\,3 Mpc are quite rare within the overall population of GRG \citep[e.g., $\sim$\,1\% in ][]{Oei.LoTSS}. Its radio structure spans from the north-west (NW) direction to the south-east (SE). The southern lobe appears to be larger than the northen lobe; this observable asymmetry between the two lobes might result from the relativistic Doppler effect. The high-resolution VLASS map detects the core and a one-sided jet toward the southeast direction. This suggests that the emission from the southern lobe is possibly being directed towards us, while that from the northern lobe is being directed away. Further dedicated deep high-resolution observation will uncover more details of the source. Additionally, the DESCaLS-DR9 image of the host galaxy shows a luminous bulge, characterized by a noticeable dust lane, as illustrated in Fig.\,\ref{fig:optical-montage}. Further deep optical photometry imaging will reveal the host galaxy properties.

\subsection{GRQ7 (J0847$+$3831)}
\begin{figure*}
    \centering
    \includegraphics[scale=0.5]{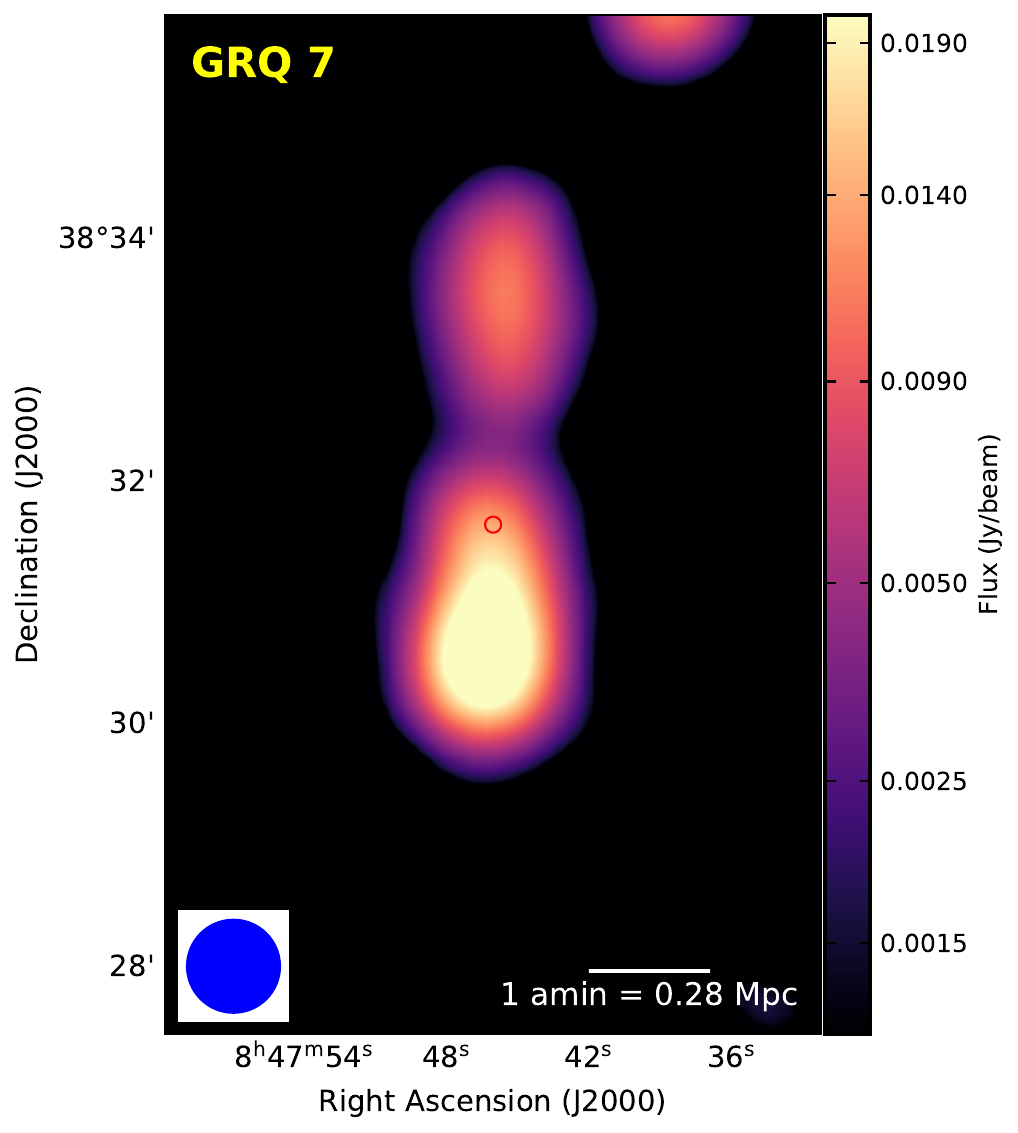}
    \includegraphics[scale=0.5]{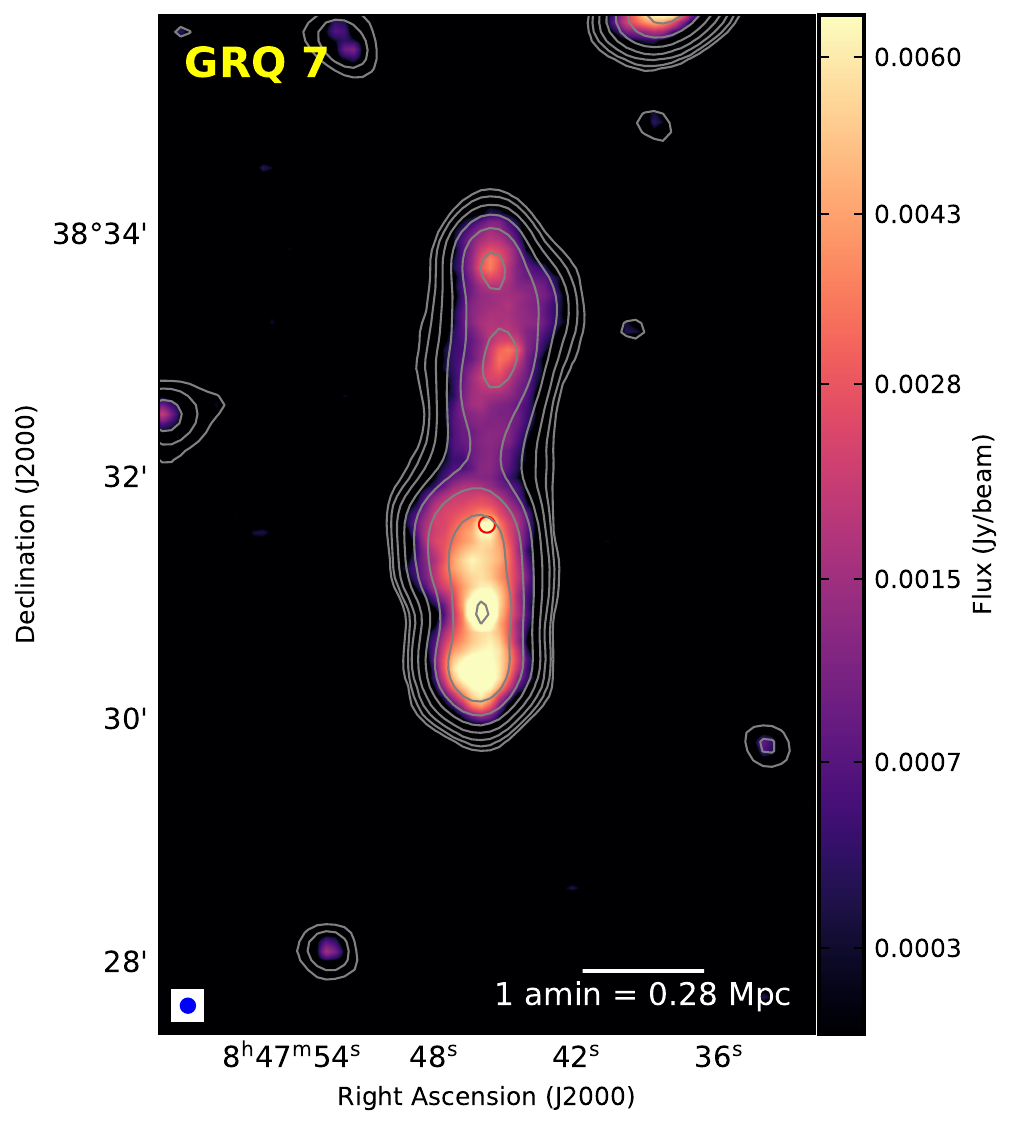}
    \includegraphics[scale=0.38]{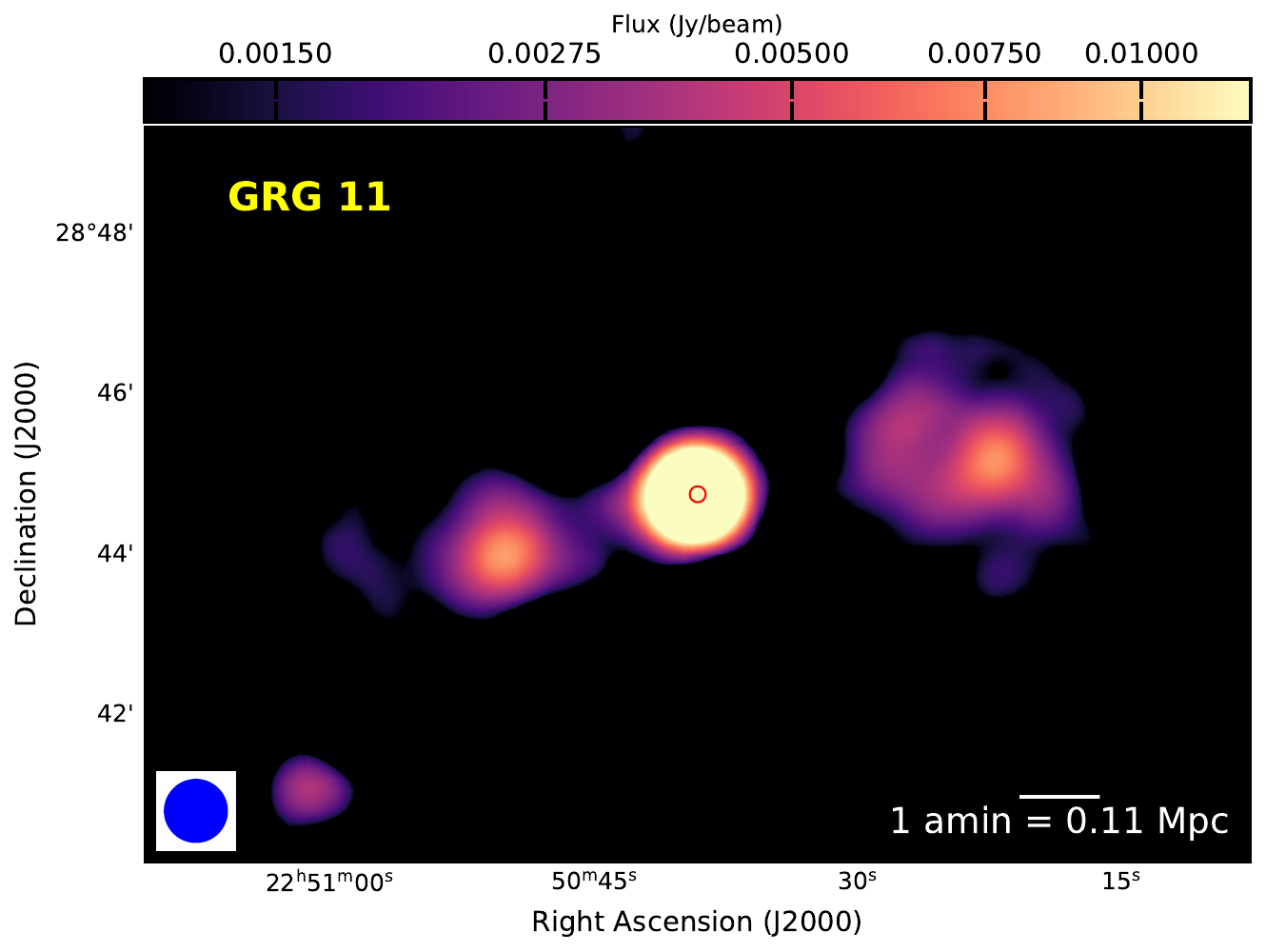}
    \includegraphics[scale=0.38]{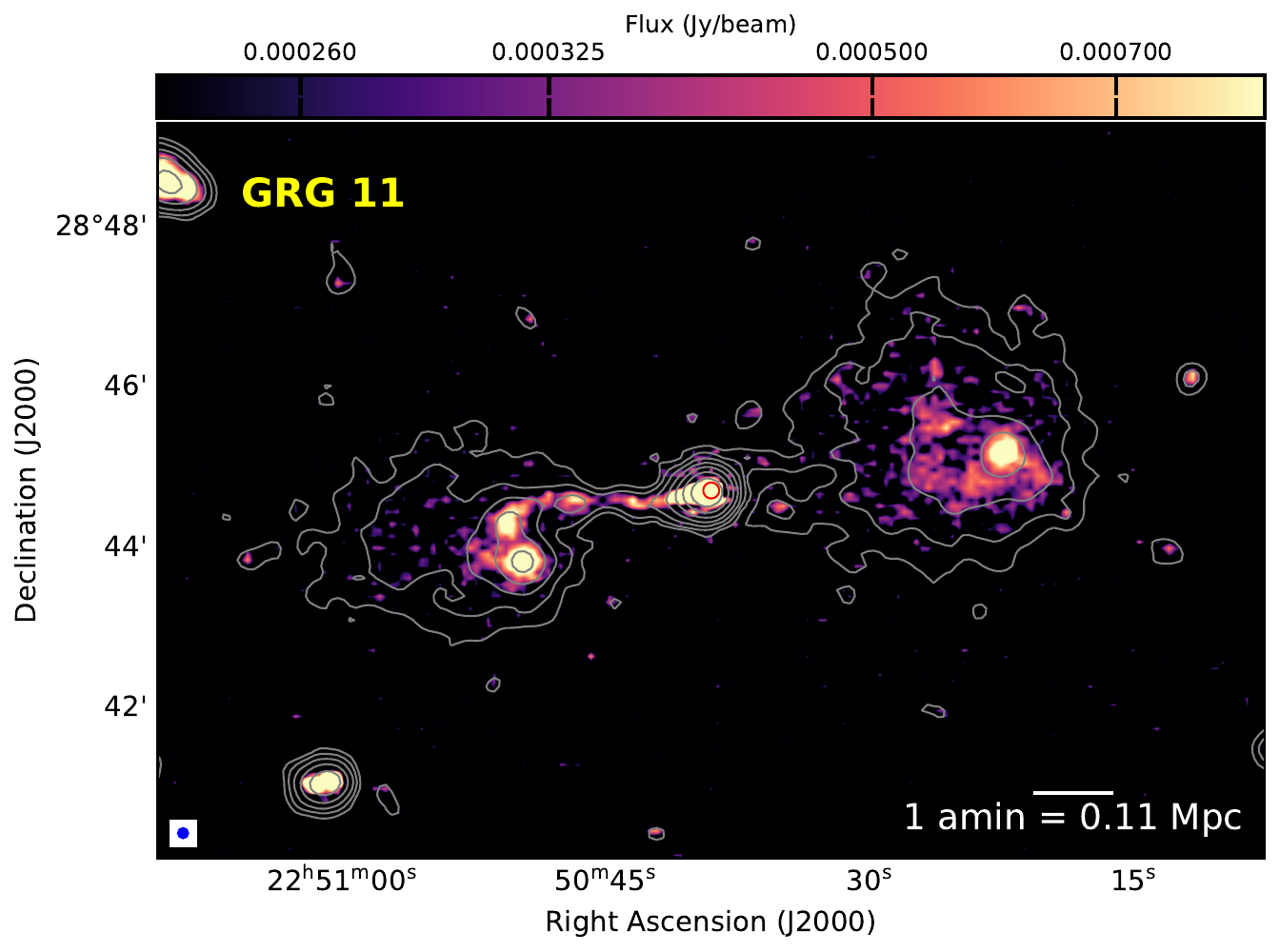}
    \caption{In the upper and lower panels we present new radio maps of GRQ7 and GRG11, respectively. On the left-hand side is the NVSS 1.4~GHz colour map with 45\arcsec\,resolution and rms of 0.45 \mjyb. On the right-hand side, LoTSS 144~MHz 6\arcsec\,colour map (rms$\sim$ 85 \ujyb) is overlaid with LoTSS 20\arcsec\,contours (rms $\sim$ 150 and 180 \ujyb ~for GRG11 and GRQ7 respectively). The contours are plotted at $\rm 3\,rms\, \times\,2^n$ (n = 0,1,2,3 ...) for both maps. The blue ellipses in the bottom left corner of the respective images indicate the beam size. The angular length scale is given in the bottom right corner of the respective figures.}
    \label{fig:GRG7_11_lofarnvss}
\end{figure*}

J0847$+$3831 is the only GRQ in our sample. GRQs compared to the GRG population are much rarer and constitute less than 30\% of the total GRG population \citep{kuzmicsethi2021,sagan3}. In the upper panel of Fig.\,\ref{fig:GRG7_11_lofarnvss}, we have presented the radio maps from NVSS and LoTSS. The overall extent of the source is well captured in the LoTSS 20\arcsec map (white contours) and the NVSS 45\arcsec map, while finer structural details are revealed in the higher-resolution LoTSS 6\arcsec map. In both maps, the radio core is not resolved; however, we can observe it in the high-resolution map of VLASS. The southern lobe appears brighter than the northern one, possibly due to an unresolved structure (e.g. knot) embedded in the lobe. The radio core coinciding with the host galaxy is closer to the southern lobe, indicating a possible effect of relativistic beaming with the southern lobe pointed towards our line of sight. We estimate the integrated spectral index of 0.83$\pm$0.02 using the 1.4~GHz NVSS and 144~MHz LoTSS maps. The 144~MHz total flux is 596$\pm$2 mJy and considering the spectral index of 0.83 the total radio power is 1.92\,$\rm \times 10^{26}$ \whz for this GRG.

\subsection{GRG8 (J1908$+$6957)}
GRG8 is a very low surface brightness RG and it spans from the northeast (NE) to the southwest (SW). Both lobes are diffuse and without bright hotspots, with the northern lobe being significantly fainter than the southern lobe. The radio core has the flux density of $\sim$\,25 mJy at 3~GHz (VLASS), making it the second brightest in our sample.  The host galaxy shows the presence of an extended disc with a luminous bulge as shown in Fig.~\ref{fig:optical-montage}. Considering the features of the lobes, core, and the host galaxy, GRG8 appears to be undergoing a restarted phase.

\subsection{GRG9 (J1930$+$2803)}
In the sky plane, the host galaxy of the GRG is situated in a densely populated area of our Milky way galaxy (galactic latitude b= 4.7\deg), evident from Fig.\,\ref{fig:montage}, which displays numerous foreground objects, possibly a star cluster. The optical brightness of the host galaxy of GRG suffers from a galactic extinction, making the photometric redshift estimation very unreliable. Moreover, identifying the host galaxy in such a crowded region is also very challenging. Consequently, dedicated spectroscopic observations were essential to accurately determine the redshift, as photometric methods were rendered unreliable due to galactic extinction and crowding. GRGs are rarely discovered or identified in this region of the sky due to its complexity, dominated by the dense emission and crowding from our parent Milky Way galaxy, making this confirmation particularly significant. Although GRG9 exhibits an FR-II-type morphology, its overall morphology is quite asymmetric, with the southern lobe being close to the host galaxy. We observe a bright unresolved core with extended radio emission, indicating the presence of jets that are not resolved on the available NVSS map of 45\arcsec. The northwest lobe (NW), which also has a longer spread, is positioned at an $\sim$\,15\deg\,offset from the core, indicating the possible influence of the surrounding local medium. 

\subsection{GRG11 (J2250$+$2844)}
The NVSS and LoTSS maps for GRG11 are shown in the lower panel of Fig.~\ref{fig:GRG7_11_lofarnvss}, where the LoTSS 144~MHz map of 6\arcsec\, clearly resolves the structure of the GRG. Notably, this high-resolution map reveals a kpc-scale jet on the eastern side of the GRG, which bends 90 degrees toward the south and eventually terminates with the eastern hotspot. The projected linear size of the eastern jet\footnote{The jet meets the criteria for jets defined in \citet[]{Jet.Review.Bridle84}{}.} is approximately 275\,kpc ($\sim$ 2.5\arcmin). There are very handful of examples of RGs that exhibit detectable one-sided or two-sided jet extending $\sim$ 100\,kpc (e.g. NGC\,315, \citealt[][]{NGC315}; NGC\,6251, \citealt[][]{NGC.6251.Perley.Bridle.84}; HB\,13, \citealt[][]{CGCG.HB13.Jaegers}; CGCG 049$-$033, \citealt[][]{CGCG.Bagchi}; J2233$+$1315, \citealt[][]{Pratik.Barbell}; J0644$+$1043, \citealt{SGRG.Sethi24}). The occurrence rate of kpc-scale jet among RGs is not high and it is even rarer in GRGs, as briefly discussed in \cite{SGRG.Sethi24}. The detection rate of kpc-scale one-sided jet is only 62\,\% for redshift $<$0.15  \citep[see e.g.][and references therein]{Jet.Mullin09}{}{}. However, the high resolution of the LoTSS map enables the detection of this one-sided kpc-scale jet in GRG11. Additionally, the 20\arcsec\,LoTSS map recovers the diffuse emission from the lobes, which is not observed in the 1.4~GHz NVSS map. The 144~MHz total flux measures 368$\pm$3 mJy, with a radio power of $\sim$\,8.84\,$\rm \times 10^{24}$ \whz.

\section{Summary and discussion}\label{sec6:summary}
The only alternative for obtaining spectroscopic redshift measurements for galaxies not covered by large spectroscopic surveys like SDSS—whether due to selection criteria or limited sky coverage—is through dedicated spectroscopic observations using available telescopes. These observations are typically performed with slit-based instruments, as access to multi-object spectrographs (MOS) or integral field units (IFUs) is often constrained by their prioritisation for large-scale survey projects or other high priority targets, limiting their availability for smaller, focused studies.

Dedicated spectroscopic observations of galaxies/AGNs are crucial for addressing the gaps left by large-scale optical spectroscopic surveys, which often miss targets due to their selection functions and limited sky coverage. These surveys typically focus on specific criteria, such as brightness or colour, potentially overlooking significant populations of galaxies/AGNs. Small-sized telescopes, like those in the 2-metre class (e.g., HCT), are particularly well-suited for this task. They offer the flexibility to conduct targeted observations of these missed objects, allowing for a more comprehensive understanding of galaxy and AGN properties. Such dedicated observations can provide detailed spectral data, revealing information about the physical conditions. 

To obtain accurate redshift measurements of 11 candidate GRGs, we conducted spectroscopic observations with the 2-m HCT in 2016. Using our wavelength-calibrated spectra, we determined precise redshifts and confirmed the giant nature of these radio sources, three of which are being reported for the first time in this paper. The projected linear sizes of GRG2 and GRG3 exceed 2.5 Mpc. GRGs with sizes greater than 2.5 Mpc are rare compared to GRGs of smaller size \citep{GRSREV2023}. We classified the host galaxies as LERGs and HERGs (NLRGs or BLRGs) based on the detection of specific optical emission lines and their widths. Although the sample is small, these spectroscopic observations provide accurate redshifts and AGN classifications, contributing to future studies of GRGs and their environments.

The predominance of FR-II morphologies in our sample, coupled with the observed CDF range, indicates that the majority of GRGs in this study are oriented at moderate angles to the line of sight. The observed asymmetries in ALR ($1.12 \leq R_{\theta} \leq 1.59$) and variations in CDF ($0.03 \leq \text{CDF} \leq 2.18$) suggest significant environmental influences on jet propagation, with possible contributions from anisotropic IGM densities and relativistic beaming effects. Additionally, the FR-II morphologies also could indicate that, at the time of their birth, the AGN jets of these GRGs were among the most powerful.

Another important aspect of this paper is the importance of spectroscopic redshift in identifying the extended low surface brightness GRGs from wide-area sky surveys. Recent radio surveys have reached deep sensitivities, which is finding large samples of sources with diffuse extended low surface brightness, which are often at higher redshifts. For example, nearly 50\% of sources have photometric redshift measurements in the large new sample ($\sim$\,2000) of GRG found from the LoTSS DR2 \citep{Oei.LoTSS} in a relatively small sky area of $\sim$\,5600 \sqdeg. Also, tens of sources were found with no optical identification owing to the possible high redshift nature of the sources. There are thousands of low surface brightness GRG waiting to be discovered from new radio sky surveys with dedicated spectroscopic observations. Identifying a larger number of giants is crucial for accurately determining the sky density of GRG and advancing our understanding of the growth and evolution of radio sources.

The recent GRG catalogue by \citet{2024Mostert}, comprising $\sim$\,11.5 thousand sources, is constructed from three main components: newly identified sources using machine learning (ML) methods, newly classified sources from the radio galaxy zoo (RGZ) project \citep{Hardcastle2023}, and previously reported sources compiled from the literature. However, a significant limitation of this catalogue is its reliance on photometric redshift measurements for a large fraction of its sources. Focusing specifically on the newly reported sources from ML and RGZ, we applied a redshift error threshold of $> 10^{-3}$ to identify photometric redshifts. For the RGZ sample, $\sim$\,77\% of the sources have photometric redshifts, while for the ML-identified sample, this fraction increases to about 83\%. These results highlight that the majority of newly reported sources in these subsets are based on photometric redshifts, with only a small fraction having spectroscopic measurements. Using the latest large spectroscopic survey from DESI \citep{DESI2024}, we could only identify spectroscopic redshifts for fewer than 100 unique sources from the combined RGZ and ML subsets of the \citet{2024Mostert} catalogue. This highlights a critical limitation: the heavy reliance on photometric redshifts compromises size accuracy and may lead to misclassification, as accurate spectroscopic data could reveal that some sources do not meet the GRG size threshold.

The upcoming data from Euclid surveys \citep{EUCLID} holds significant promise for advancing GRG research by providing spectroscopic measurements for galaxies in the redshift range $0.8 \lesssim z \lesssim 2$. This is a critical redshift window where many GRGs currently have only photometric redshifts or lack redshift measurements altogether, and in some cases, even their host galaxies remain unidentified. Euclid's ability to obtain spectroscopic data over large sky areas will confirm high-redshift GRG candidates, refine size estimates and expand the GRG sample within this redshift range. Euclid's data will enable robust studies of GRG evolution across cosmic epochs, revealing insights into their growth mechanisms, environmental influences, and AGN feedback. By filling critical gaps in redshift data, Euclid data is expected to advance our understanding of GRGs and their role in the cosmic web.

\begin{acknowledgements}
We thank the anonymous referee for her/his valuable comments and suggestions. We thank the staff at IAO, Hanle, and CREST, Hosakote, operated by the Indian Institute of Astrophysics, Bengaluru (India), for their support in facilitating these observations. PD acknowledges support from the Indo-French Centre for the Promotion of Advanced Research (CEFIPRA) under project no. 5204-2, active between 2015 and 2018, during which the observations for this study were carried out as well as the IUCAA, Pune (India).

S.S. and M.J. were partly supported by the Polish National Science Centre (NCN) grant UMO-2018/29/B/ST9/01793. S.S. also acknowledges the Jagiellonian University grants: 2022-VMM U1U/272/NO/10, 2022-SDEM U1U/272/NO/15, and VMM-2023 U1U/272/NO/10. 

We acknowledge that this work has made use of \textsc{ipython} \citep{per07}, \textsc{astropy} \citep{astropy}, \textsc{matplotlib} \citep{plt}, \textsc{SciPy} \citep{SciPy.2020}.
\end{acknowledgements}
\bibliographystyle{aa} 
\bibliography{ref} 

\begin{appendix}
\onecolumn
\section{Spectral line and redshift identification table}\label{sec:app_line}
\begin{center}
\setlength{\tabcolsep}{2.5pt} 
\renewcommand{\arraystretch}{1.00} 
\begin{longtable}{clccccc}
\captionsetup{width=\textwidth}
\caption{Spectral line and redshift identification. Col(1): Object name, Col(2): Observed lines with rest frame wavelength in \AA, Col(3): Observed wavelength in \AA, Col(4): Redshift, Col(5): Error in redshift, Col(6): Average redshift with average error in redshift and Col(7): AGN type.}\\
\hline 
\multicolumn{1}{c}{Name} &\multicolumn{1}{c}{Line} &\multicolumn{1}{c}{$\rm \lambda_{obs}$ ($\AA$)} &\multicolumn{1}{c}{$z$} &\multicolumn{1}{c}{$z_{\rm error}$} &\multicolumn{1}{c}{$z_{\rm avg}$} &AGN \\
&&&&& &\multicolumn{1}{c}{Type}\\
\multicolumn{1}{c}{(1)} &\multicolumn{1}{c}{(2)} &\multicolumn{1}{c}{(3)} &\multicolumn{1}{c}{(4)} &\multicolumn{1}{c}{(5)} &\multicolumn{1}{c}{(6)}&\multicolumn{1}{c}{(7)}\\
\hline 
\endfirsthead
\multicolumn{3}{c}%
{{\bfseries \tablename\ \thetable{} -- continued from previous page}} \\
\hline 
\multicolumn{1}{c}{Name} &\multicolumn{1}{c}{Line} &\multicolumn{1}{c}{$\rm \lambda_{obs}$ ($\AA$)} &\multicolumn{1}{c}{$z$} &\multicolumn{1}{c}{$z_{\rm error}$} &\multicolumn{1}{c}{$z_{\rm avg}$} &AGN \\
&&&&& &\multicolumn{1}{c}{Type}\\
\multicolumn{1}{c}{(1)} &\multicolumn{1}{c}{(2} &\multicolumn{1}{c}{(3)} &\multicolumn{1}{c}{(4)} &\multicolumn{1}{c}{(5)} &\multicolumn{1}{c}{(6)} &\multicolumn{1}{c}{(7)}\\
\hline
\endhead
\hline \multicolumn{7}{r}{\textit{Continued on next page}}
\endfoot
\endlastfoot
& H11 (3771) & 4595.46 & 0.21875 & 0.00259 &  & \\
& H9 (3835) & 4662.42 & 0.21563 & 0.00247 &  & \\
& [O\,{\sc iii}] 1(4363) & 5324.68 & 0.22036 & 0.00287 & & \\
J0125+0703 & $\rm H_{\beta}$ (4861) & 5935.98 & 0.22106 & 0.00492 & 0.21984$\pm$0.00277  & LERG \\
& [O\,{\sc iii}] 2(4959) & 6066.84 & 0.22342 & 0.00126 &  & \\
&  [O\,{\sc iii}] 3(5007) & 6107.41 & 0.21981 & 0.00250 &  & \\
\hline

& $\rm H_{\beta}$ (4861) & 6160.97 & 0.26734 & 0.00241 &  & \\
&  [O\,{\sc iii}] 2(4959) & 6284.22 & 0.26726 & 0.00257 &  & \\
J0151-1112 &  [O\,{\sc iii}] 3(5007) & 6345.15 & 0.26730 & 0.00272 & 0.26740$\pm$0.00219 & NLRG \\
& $\rm H_{\alpha}$ (6563)  & 8318.03 & 0.26745 & 0.00253 &  & \\
& [N\,{\sc ii}] 3(6583) & 8346.51 & 0.26781 & 0.00152 &  & \\
& [S\,{\sc ii}] 1(6716) & 8511.22 & 0.26722 & 0.00138 &  & \\
\hline

& [Ne\,{\sc iii}](3967) & 4744.10 & 0.19577 & 0.00387 &  & \\
& $\rm H_{\beta}$ (4861) & 5822.90 & 0.19780 & 0.00096 & & \\
J0235+1011 & [O\,{\sc iii}] 2(4959) & 5945.32 & 0.19891 & 0.00134 & 0.19980$\pm$0.00185 & LERG  \\
& [O\,{\sc iii}] 3(5007) & 5994.52 & 0.19727 & 0.00120 &  & \\
& [N\,{\sc ii}] 1(5755) & 6958.91 & 0.20927 & 0.00189 &  & \\
\hline

& H12 (3750) & 5279.13 & 0.40771 & 0.00476 &  & \\
& $\rm H_{\delta}$ (4102) & 5793.72 & 0.41250 & 0.00242 &  & \\
& $\rm H_{\beta}$ (4861) & 6850.12 & 0.40910 & 0.00235 &  & \\
& [O\,{\sc iii}] 2(4959) & 6994.09 & 0.41041 & 0.00528 &  & \\
J0429+0033 &  [O\,{\sc iii}] 3(5007) & 7063.45 & 0.41076 & 0.00646 & 0.41134$\pm$0.00414 & NLRG \\
& [O\,{\sc i}] (6300) & 8916.38 & 0.41523 & 0.00309 &  & \\
& $\rm H_{\alpha}$ (6563) & 9255.72 & 0.41033 & 0.00503 &  & \\
&  [N\,{\sc ii}] 3(6583) & 9313.41 & 0.41468 & 0.00372 &  & \\
\hline

& $\rm H_{\beta}$ (4861) & 5811.16 & 0.19539 & 0.00124 &  & \\
&  [O\,{\sc iii}] 2(4959) & 5938.56 & 0.19755 & 0.00170 &  & \\
&  [O\,{\sc iii}] 3(5007) & 5978.47 & 0.19406 & 0.00104 &  & \\
J0754+2324 & [N\,{\sc ii}] 2(6548) & 7830.26 & 0.19582 & 0.00130 & 0.19622$\pm$0.00134  & LERG \\
& $\rm H_{\alpha}$(6563) & 7851.23 & 0.19632 & 0.00099 &  & \\
& [N\,{\sc ii}] 3(6583) & 7888.18 & 0.19819 & 0.00179 &  & \\\hline

& $\rm H_{\epsilon}$ (3970)  & 4937.31 & 0.24363 & 0.00112 &  & \\
& $\rm H_{\delta}$ (4102) & 5060.90 & 0.23384 & 0.00235 &  & \\
J0824+0140 & $\rm H_{\beta}$ (4861) & 6055.55 & 0.24566 & 0.00164 & 0.24269$\pm$0.00168 &  NLRG \\
&  [O\,{\sc iii}] 2(4959) & 6182.68 & 0.24678 & 0.00166 &  & \\
&  [O\,{\sc iii}] 3(5007) & 6226.16 & 0.24353 & 0.00163 &  & \\
\hline

&  [O\,{\sc iii}] 1(4363) & 5730.74 & 0.31342 & 0.00147 &  & \\
& $\rm H_{\beta}$ (4861) & 6383.61 & 0.31314 & 0.00100 &  & \\
J0847+3831 &  [O\,{\sc iii}] 2(4959) & 6512.75 & 0.31334 & 0.00213 & 0.31342$\pm$0.00696 & Quasar \\
&  [O\,{\sc iii}] 3(5007) & 6575.53 & 0.31331 & 0.00249 &  & \\
& $\rm H_{\alpha}$(6563) & 8622.85 & 0.31390 & 0.02768 &  & \\
\hline

&  [O\,{\sc iii}] 1(4363) & 5308.92 & 0.21675 & 0.00371 &  & \\
& $\rm H_{\beta}$ (4861)  & 5929.03 & 0.21963 & 0.00300 &  & \\
J1908+6957 &  [O\,{\sc iii}] 2(4959) & 6035.31 & 0.21706 & 0.00045 & 0.21755$\pm$0.00162 & LERG \\
&  [O\,{\sc iii}] 3(5007) & 6081.47 & 0.21463 & 0.00040 &  & \\
& [N\,{\sc ii}] 1(5755) & 7004.92 & 0.21727 & 0.00093 &  & \\
& HeI (5876) & 7168.12 & 0.21997 & 0.00120 &  & \\\hline

&  [O\,{\sc iii}] 3(5007) & 5473.82 & 0.09327 & 0.00211 &  & \\
& [N\,{\sc ii}] 1(5755) & 6295.27 & 0.09395 & 0.00030 &  & \\
& [N\,{\sc ii}] 2(6548) & 7156.72 & 0.09296 & 0.00055 &  & \\
J1930+2803 & $\rm H_{\alpha}$ (6563) & 7173.26 & 0.09302 & 0.00101 & 0.09323$\pm$0.00116  & NLRG \\
& [N\,{\sc ii}] 3(6583) & 7196.83 & 0.09318 & 0.00170 &  & \\
& [S\,{\sc ii}] 1(6716) & 7338.38 & 0.09259 & 0.00152 &  & \\
& [S\,{\sc ii}] 2(6731) & 7361.33 & 0.09367 & 0.00092 &  & \\
\hline

& $\rm H_{\beta}$ (4861) & 5337.81 & 0.09802 & 0.00067 &  & \\
&  [O\,{\sc iii}] 3(5007) & 5469.40 & 0.09239 & 0.00059 &  & \\
J2059+2434 & [N\,{\sc ii}] 1(5755) & 6292.51 & 0.09347 & 0.00177 & 0.09372$\pm$0.00138  & LERG \\
& $\rm H_{\alpha}$ (6563) & 7176.98 & 0.09358 & 0.00304 &  & \\
& [N\,{\sc ii}] 3(6583) & 7194.24 & 0.09278 & 0.00067 &  & \\
& [S\,{\sc ii}] 1(6716) & 7334.77 & 0.09206 & 0.00154 &  & \\\hline

&  [O\,{\sc iii}] 3(5007) & 5529.15 & 0.10432 & 0.00138 &  & \\
& [N\,{\sc ii}] 1(5755) & 6325.85 & 0.09926 & 0.00290 & & \\
J2250+2844 & [O\,{\sc i}] (6300) & 6903.30 & 0.09571 & 0.00202 & 0.09676$\pm$0.00193  & LERG \\
& $\rm H_{\alpha}$ (6563) & 7174.89 & 0.09327 & 0.00158 & & \\
& [N\,{\sc ii}] 3(6583) & 7210.99 & 0.09533 & 0.00158 & & \\
& [S\,{\sc ii}] 1(6716) & 7338.85 & 0.09266 & 0.00209 & & \\
\hline
\label{tab:2_spectra_lines}
\end{longtable}
\end{center}

\section{Wavelength-calibrated HCT optical spectra}\label{sec:app}
\begin{figure*}[h]
    \includegraphics[scale=0.34]{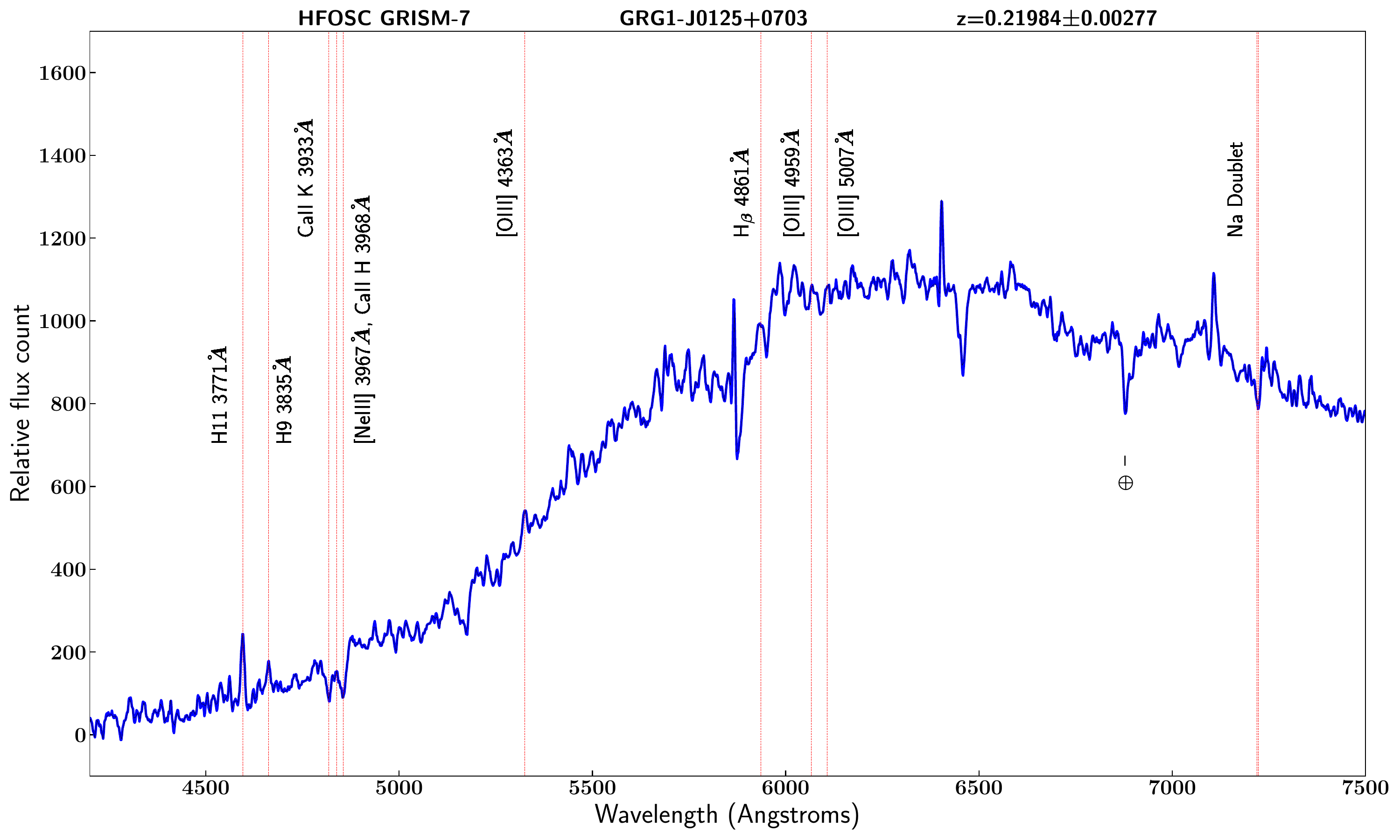}
    \caption{HCT Optical spectra of GRG1: J0125+0703}
    \label{fig:0125}
\end{figure*}

\begin{figure*}
    \includegraphics[scale=0.34]{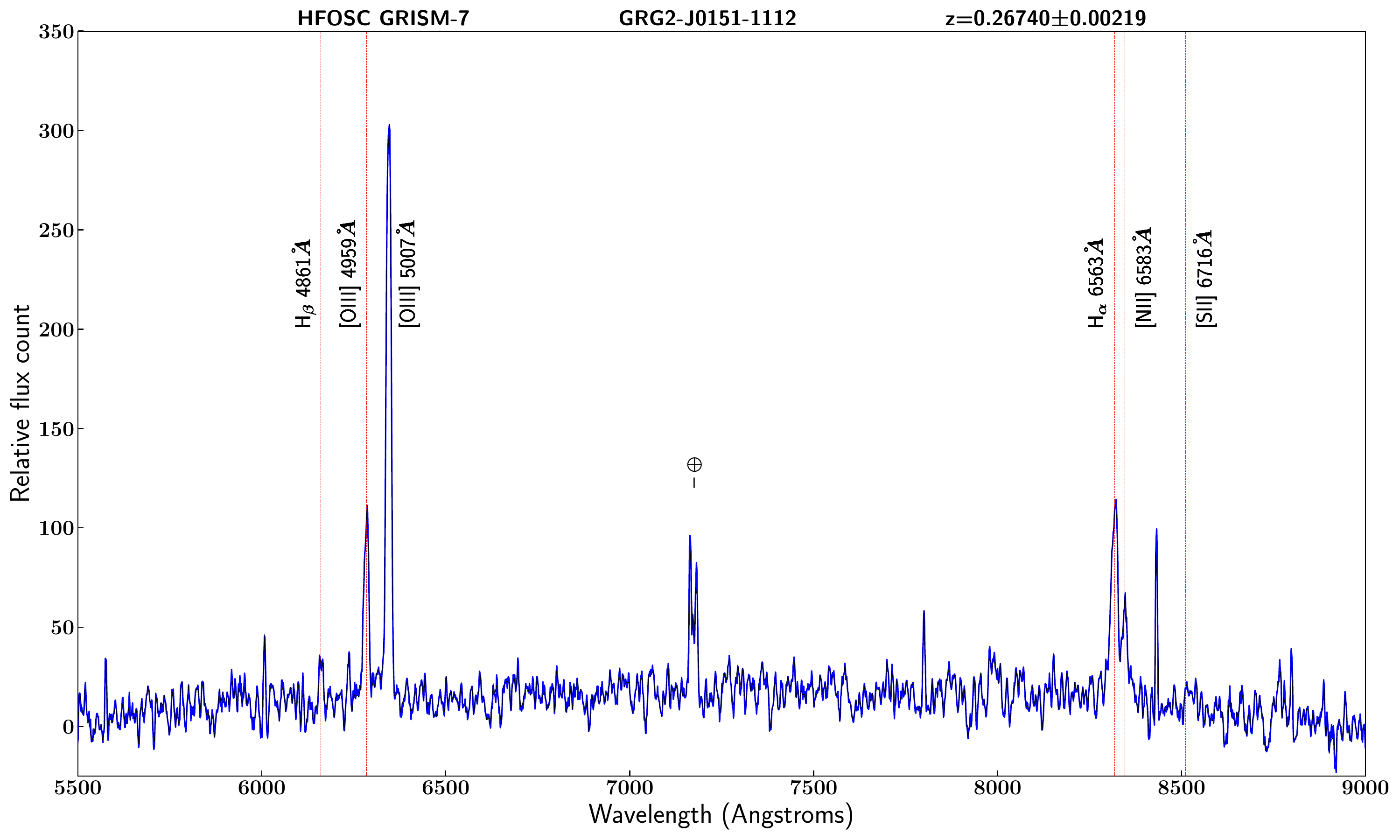}
    \caption{HCT Optical spectra of GRG2: J0151-1112}
    \label{fig:0151}
\end{figure*}

\begin{figure*}
    \includegraphics[scale=0.34]{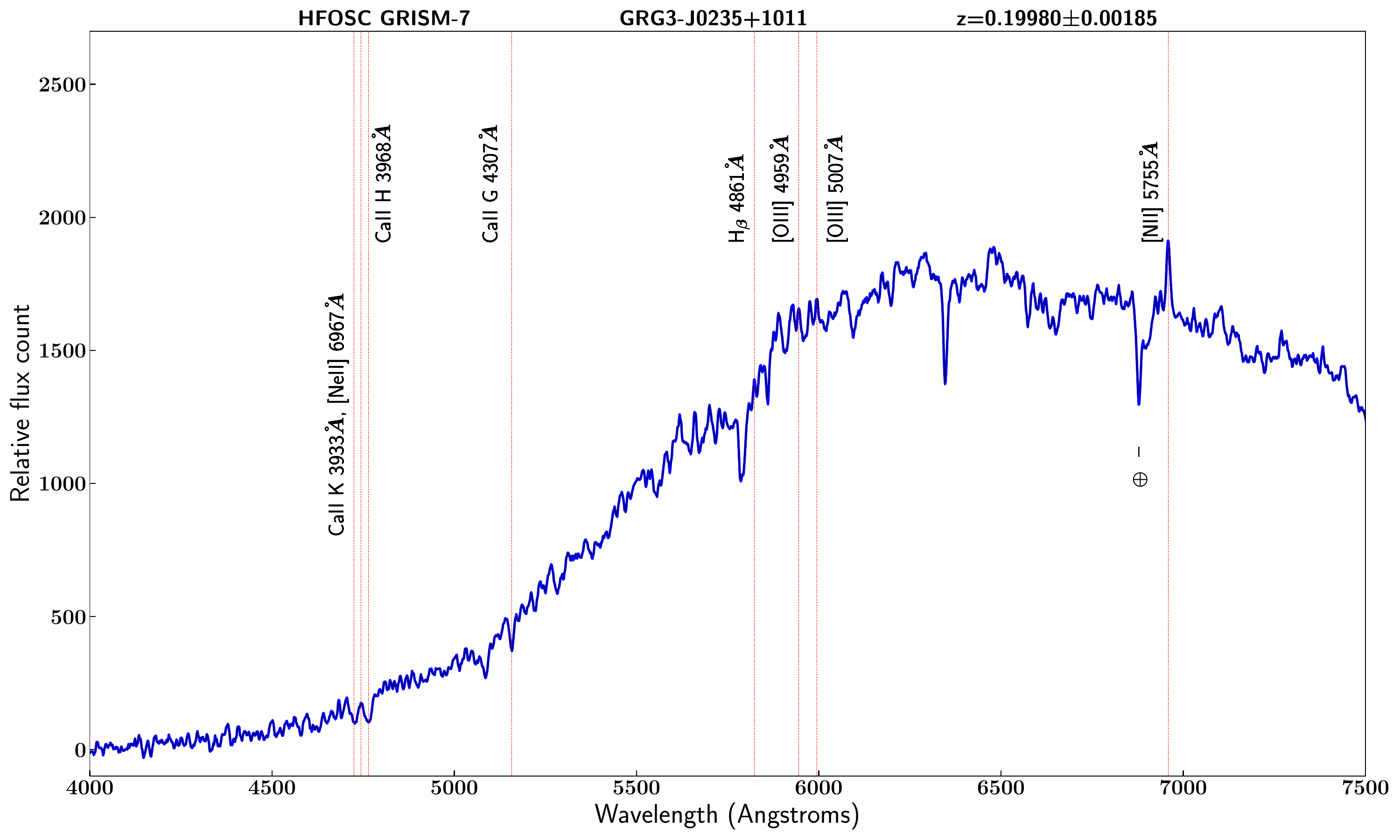}
    \caption{HCT Optical spectra of GRG3: J0235+1011}
    \label{fig:0235}
\end{figure*}

\begin{figure*}
    \includegraphics[scale=0.34]{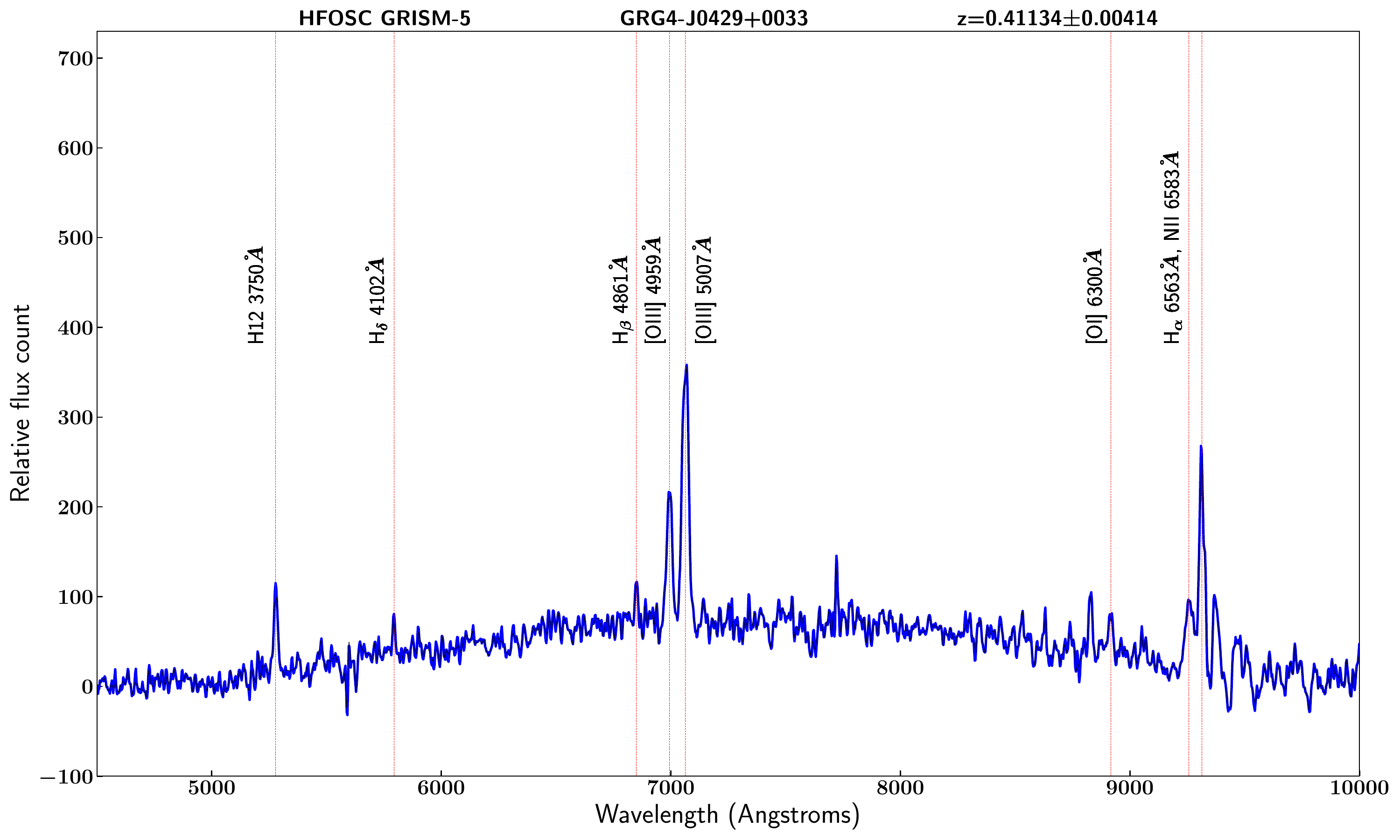}
    \caption{HCT Optical spectra of GRG4: J0429+0033}
    \label{fig:0429}
\end{figure*}

\begin{figure*}
    \includegraphics[scale=0.34]{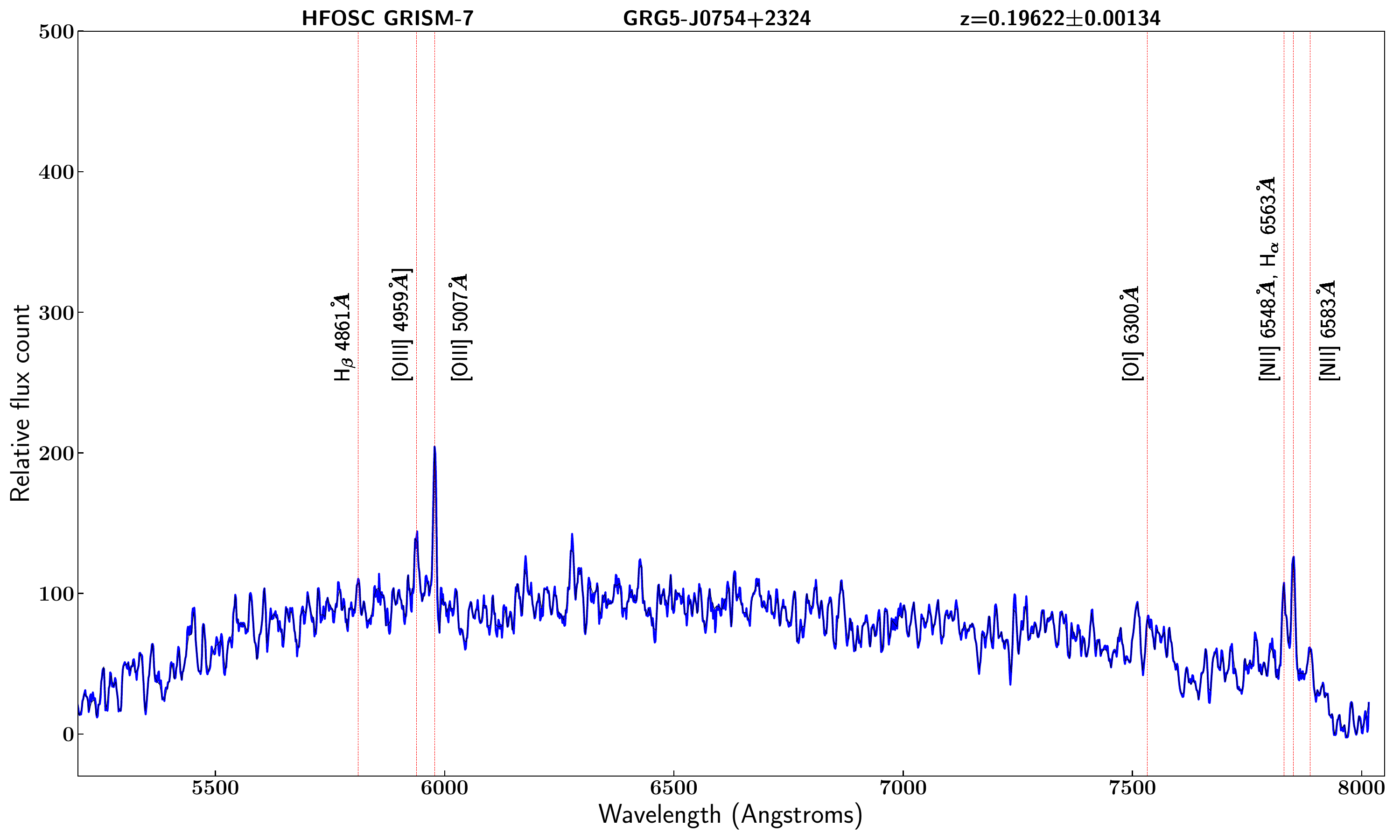}
    \caption{HCT Optical spectra of GRG5: J0754+2324}
    \label{fig:0754}
\end{figure*}

\begin{figure*}
    \includegraphics[scale=0.34]{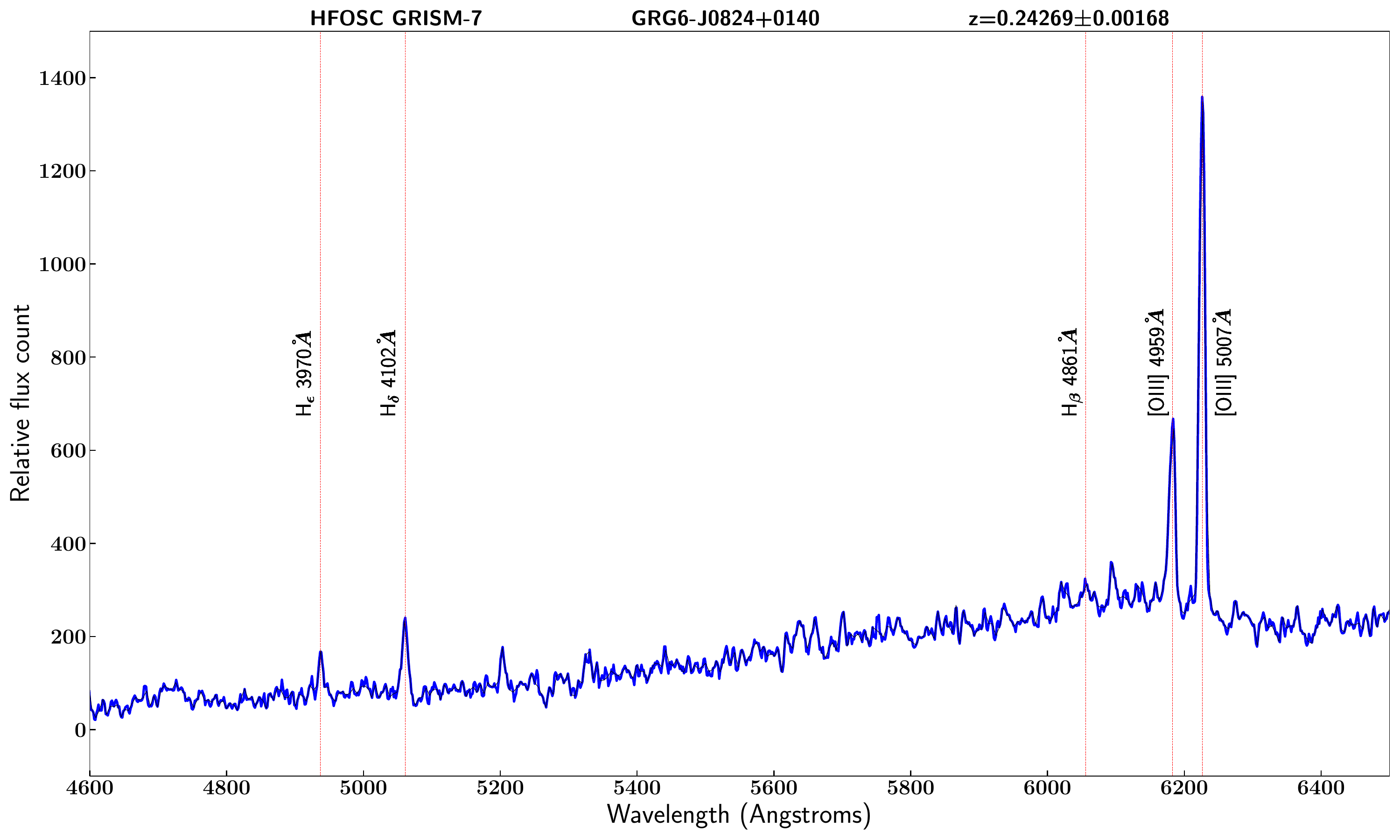}
    \caption{HCT Optical spectra of GRG6:J0824+0140}
    \label{fig:0824}
\end{figure*}

\begin{figure*}
    \includegraphics[scale=0.34]{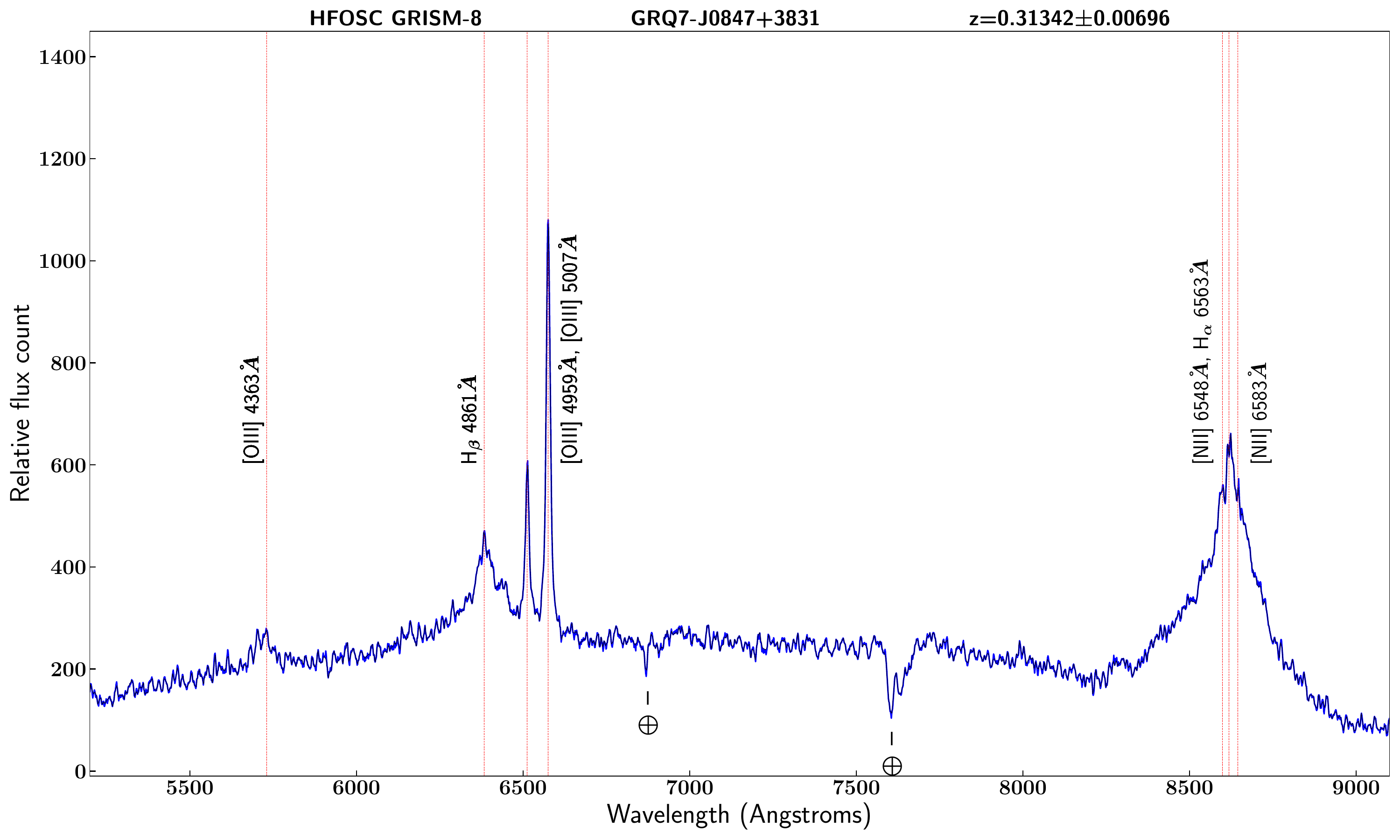}
    \caption{HCT Optical spectra of GRQ7: J0847+3831}
    \label{fig:0847}
\end{figure*}

\begin{figure*}
    \includegraphics[scale=0.34]{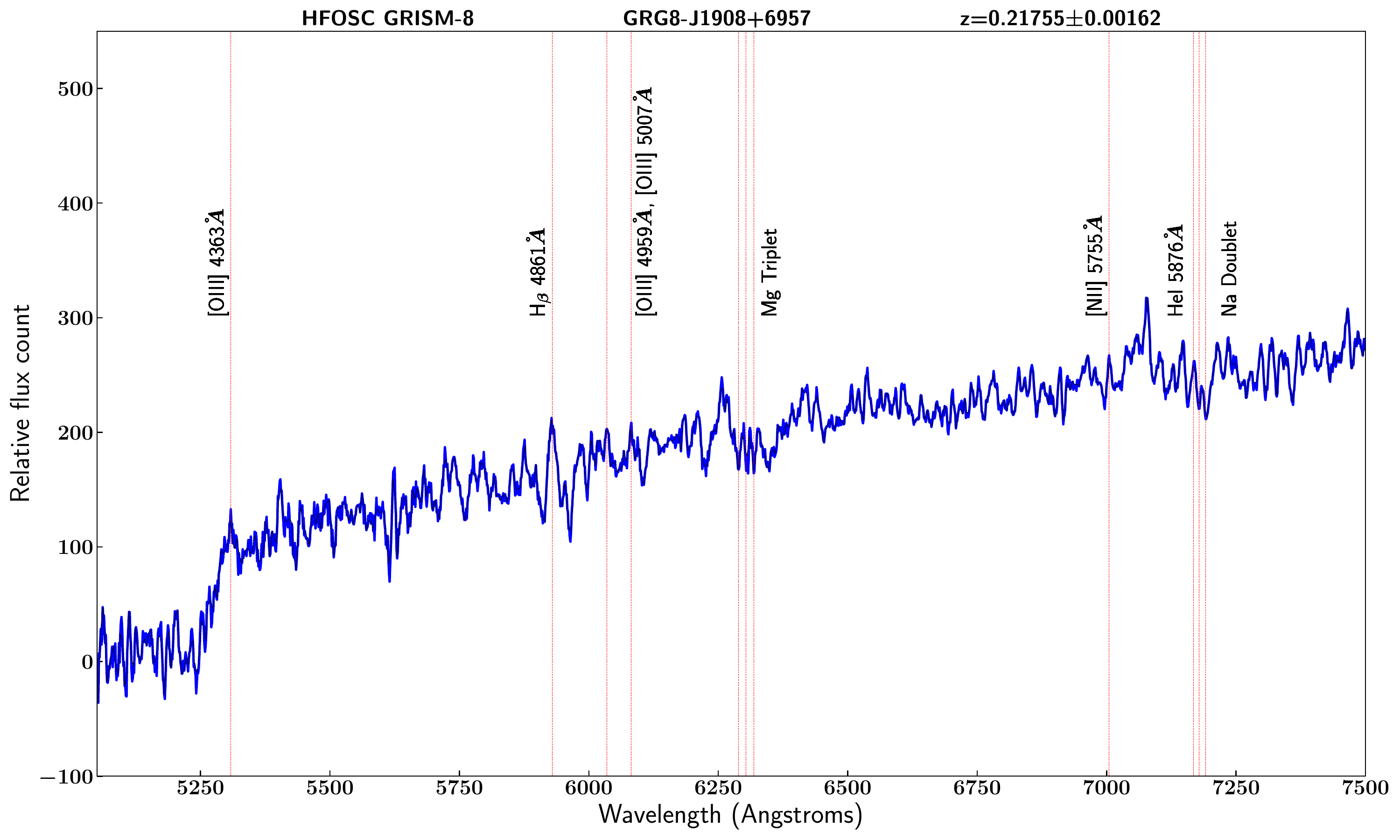}
    \caption{HCT Optical spectra of GRG8: J1908+6957}
    \label{fig:1908}
\end{figure*}

\begin{figure*}
    \includegraphics[scale=0.34]{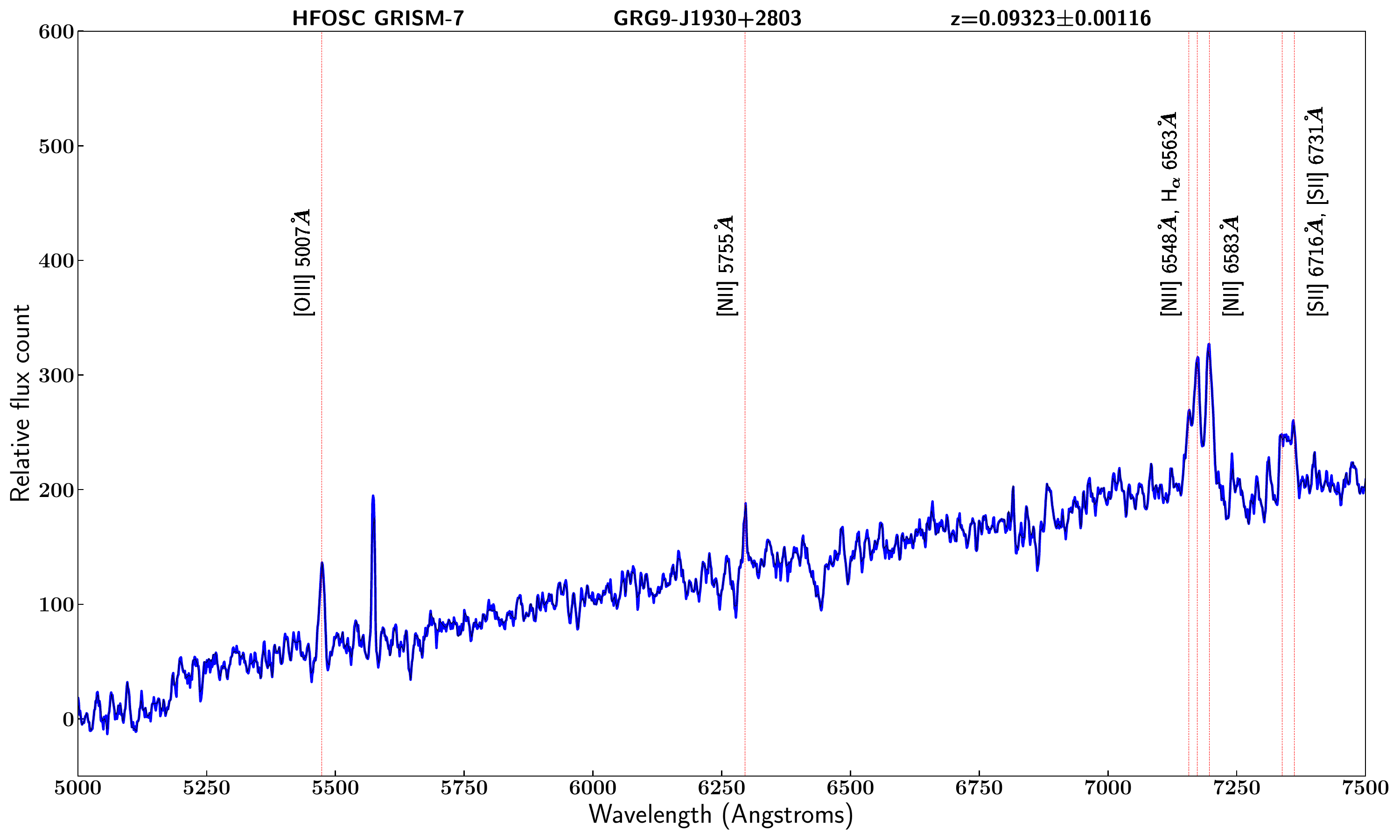}
    \caption{HCT Optical spectra of GRG9: J1930+2803}
    \label{fig:1930}
\end{figure*}

\begin{figure*}
    \includegraphics[scale=0.34]{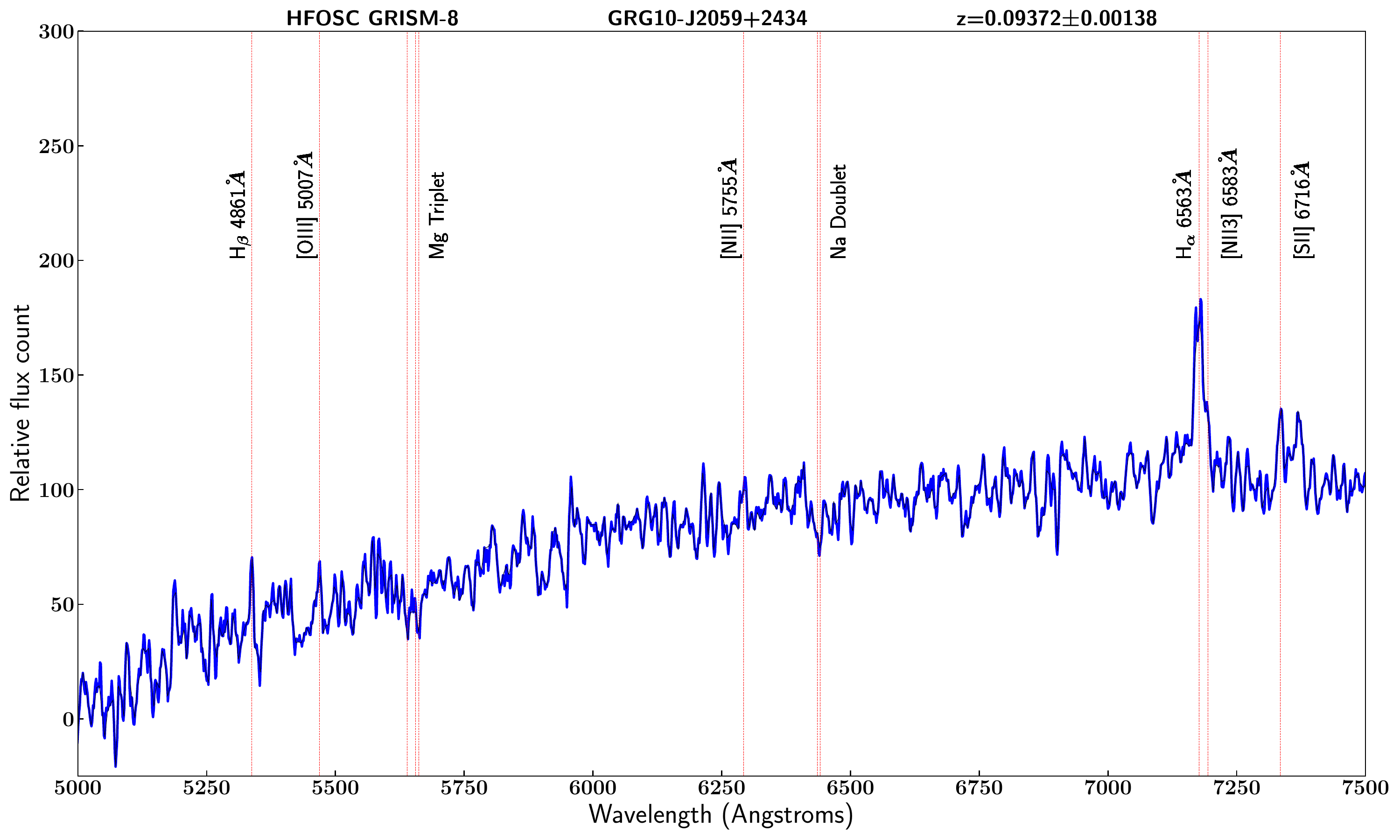}
    \caption{HCT Optical spectra of GRG10: J2059+2434}
    \label{fig:2059}
\end{figure*}

\begin{figure*}
    \includegraphics[scale=0.34]{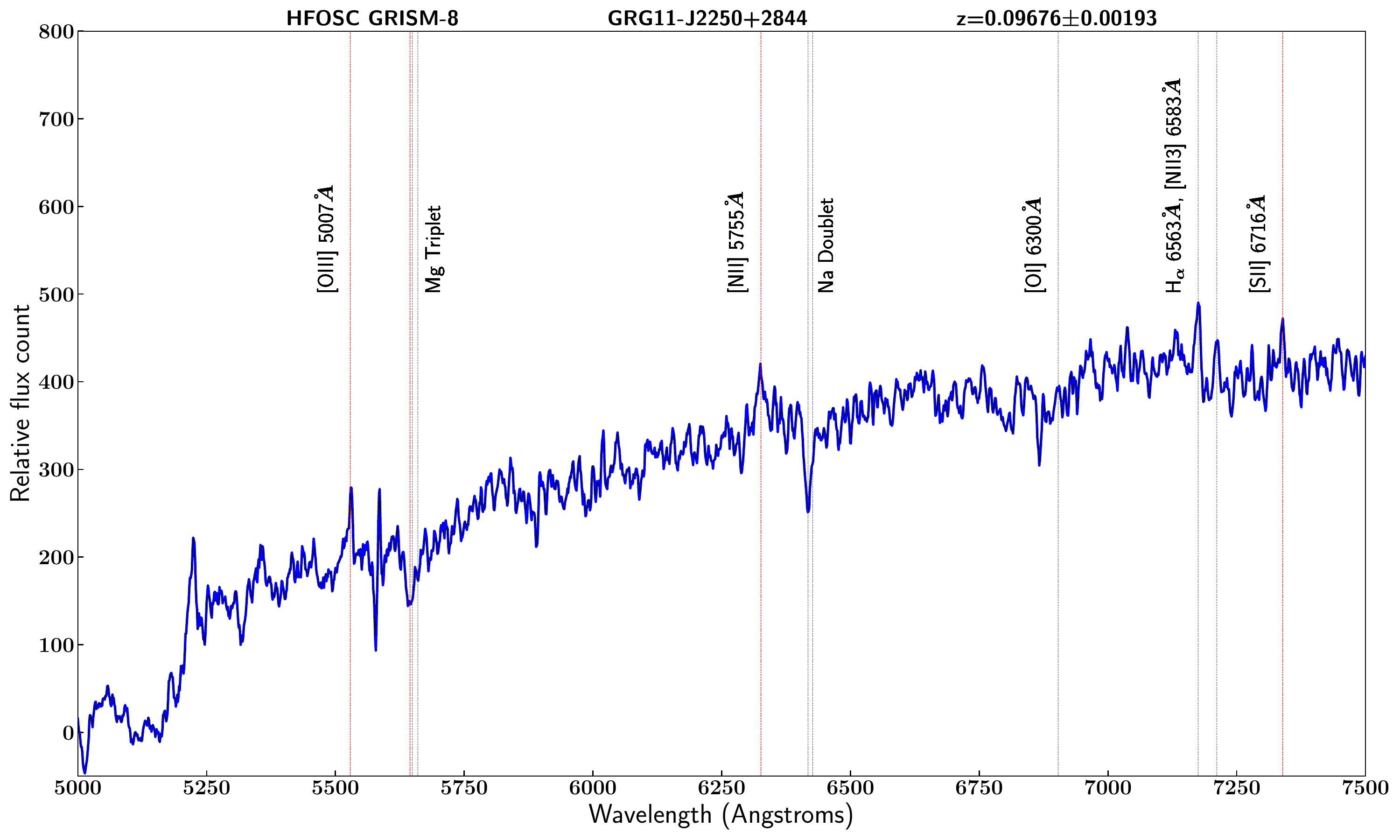}
    \caption{HCT Optical spectra of GRG11: J2250+2844}
    \label{fig:2250}
\end{figure*}

Below, we present the wavelength-calibrated spectra of all eleven GRGs in our sample, shown in the observed frame (refer to Figs. \ref{fig:0125} to \ref{fig:2250}).

\end{appendix}

\end{document}